# 3D diffusion MRI using simultaneous multi-slab with blipped-CAIPI (blipped-SMSlab) in a 4D k-space framework

Simin Liu[1], Jieying Zhang[1], Diwei Shi[2], Hua Guo[1*]

[1]Center for Biomedical Imaging Research, Department of Biomedical Engineering, School of Medicine, Tsinghua University, Beijing, China

[2]Center for Nano & Micro Mechanics, Department of Engineering Mechanics, Tsinghua University, Beijing, China

**Running title:** 3D dMRI with blipped-SMSlab

**Word Count:** ~5000

**\*Correspondence to:**

Hua Guo, PhD

Center for Biomedical Imaging Research

Department of Biomedical Engineering

Tsinghua University, Beijing, China

Phone: +86-010-6279-5886

Email: huaguo@tsinghua.edu.cn

**Grant sponsor:**

This work was supported by Beijing Municipal Natural Science Foundation (L192006) and the National Natural Science Foundation of China (61971258).

**Symbols**

$R_{mb}$: multi-band factor

$k_z$ / $k_m$: the intra-slab / inter-slab encoding dimension of the 4D k-space

$z_{\text{gap}}$: the distance between the center of the simultaneously excited slabs

$\varphi_{\text{ramp}}$: the $k_m$ gradient-induced intra-slab linear phase ramp

$\varphi_{\text{gap}}$: the $k_z$ gradient-induced inter-slab phase offset

$\varphi_{\text{off\_iso}}$: the $k_m$ / $k_z$ gradient-induced off-isocenter phase error

**Abstract**

**Purpose:** To develop an efficient simultaneous multi-slab imaging method with blipped-controlled aliasing in parallel imaging (blipped-SMSlab) in a 4D k-space framework, and demonstrate its efficacy in high-resolution diffusion MRI (dMRI).

**Theory and Methods:** First, the SMSlab 4D k-space signal expression is formulated, and the phase interferences from intra-slab and inter-slab encodings on the same physical z axis are analyzed. Then, the blipped-SMSlab dMRI sequence is designed, with blipped-CAIPI gradients for inter-slab encoding, and a 2D multi-band accelerated navigator for inter-kz-shot phase correction. Third, strategies are developed to remove the phase interferences, by RF phase modulation and/or phase correction during reconstruction, thus decoupling intra-slab and inter-slab encodings that are otherwise entangled. In vivo experiments are performed to validate the blipped-SMSlab method, and preliminarily evaluate its performance in high-resolution dMRI compared with traditional 2D imaging.

**Results:** In the 4D k-space framework, inter-slab and intra-slab phase interferences of blipped-SMSlab are successfully removed using the proposed strategies. Compared with non-CAIPI sampling, the blipped-SMSlab acquisition reduces the g-factor and g-factor-related SNR penalty by about 12%. In addition, in vivo experiments show the SNR advantage of blipped-SMSlab dMRI over traditional 2D dMRI for 1.3 and 1.0 mm isotropic resolution imaging with matched acquisition time.

**Conclusion:** Removing inter-slab and intra-slab phase interferences enables SMSlab dMRI with blipped-CAIPI in a 4D k-space framework. The proposed blipped-SMSlab dMRI is demonstrated to be more SNR-efficient than 2D dMRI and thus capable of high-quality, high-resolution fiber orientation detection.

# 1 Introduction

Diffusion MRI (dMRI) has been a crucial tool in clinical and neuroscientific applications [1-3]. At high spatial resolution, dMRI enables the mapping of fine and complex microstructure in both the white matter [4,5] and the gray matter [6,7], and the detection of small lesions in the brain [3].

Nonetheless, limited SNR and the consequent prolonged scan time pose a great challenge for achieving high spatial resolution in dMRI. Therefore, improving the SNR efficiency is essential, which is maximized in spin echo (SE) sequences when TR is equal to 1.2*T1 (about 1~2s for white matter and gray matter at 3T [8,9]). Many techniques have been developed to address this challenge in SE-EPI-based dMRI sequences, such as 2D simultaneous multi-slice (SMS) [10-14], 3D multi-slab [8,9,15], or generalized slice dithered enhanced resolution with simultaneous multi-slice (gSlider-SMS) [16]. In 2D SMS-EPI, several slices are simultaneously excited, which reduces the minimum TR by several times. However, 2D SMS-EPI may not be able to achieve SNR-efficient TRs for high-isotropic-resolution dMRI with whole-brain coverage (>100 slices), especially if parallel imaging [17,18] is also used, which reduces EPI distortions but constrains the multi-band factor. 3D multi-slab [8,9,15] and gSlider-SMS [16] encode thick slabs with $k_z$ gradients or RF pulses along the slice direction, respectively. The resultant TR ranges between the TRs of 2D EPI and single-slab 3D EPI [19], which not only improves the SNR efficiency compared with 2D EPI, but also ensures sufficient T1 recovery compared with single-slab 3D EPI [8,9]. For whole-brain dMRI with spatial resolution higher than 1.1 mm isotropic, gSlider-SMS [16,20,21] and typical 3D multi-slab acquisitions [8,22] have similar SNR efficiency (with TR approximately ranging between 3~5s). Their SNR efficiency has the potential to be further improved using thicker slabs at the expense of more RF or $k_z$ encodings. Another recent advance along this line of research is the simultaneous multi-slab (SMSlab) method, which combines SMS and 3D multi-slab techniques to achieve the TR for maximized SNR efficiency [9,23-25].

The T1 recovery-related spin-history effect for short TRs should be carefully handled, since it causes slab boundary artifacts for 3D multi-slab, SMSlab and gSlider-SMS. With newly developed slab boundary artifact correction methods, such as nonlinear inversion for slab profile encoding (NPEN) [26], revised NPEN [27] and convolutional-neural-network-enabled inversion for slab profile encoding (CPEN) [28], multi-slab and SMSlab dMRI using TR=1~2 have been achieved with slab boundary artifacts effectively suppressed [25,26,28]. Compared with multi-slab, SMSlab has the potential for higher acceleration factors due to the larger spatial variation of coil sensitivities along the slice direction [25].

In 2D SMS-EPI, blipped-controlled aliasing in parallel imaging (blipped-CAIPI) [10] effectively reduces g-factor penalty, by introducing FOV shift among simultaneously excited slices to better utilize the spatial variation of coil sensitivities. If treating blipped-CAIPI gradients as the 2$^{nd}$ group of phase encoding gradients, which are applied to the slice direction during EPI readout, 2D SMS-EPI can be described by a 3D k-space framework [29] with an interleaved under-sampling pattern along $k_y$ and the multi-band dimension. The blipped-CAIPI sampling has been successfully applied to single-slab 3D EPI fMRI [30] along $k_y$-$k_z$ to improve scan efficiency and minimize g-factor penalty. The combination of blipped-CAIPI and SMSlab, however, is prohibitively challenging, as analyzed below.

Among the SMSlab encoding strategies, the basic method treats the thickness of each slab as the FOV for intra-slab encoding [9,23,24]. In these studies, blipped-CAIPI was not used as in 2D SMS-EPI, since the blipped-CAIPI gradients are also applied along the slice direction, i.e., the inter-slab and intra-slab encodings share the same physical z axis. This brings the coupling effect between the two encodings, i.e., they affect each other as they have different requirements on the encoding gradients due to different definitions of FOV. To avoid the gradient coupling effect, Zahneisen et al proposed a 4D k-space concept and used RF encoding rather than blipped-CAIPI gradients for inter-slab encoding ($k_m$) [31]. The scan efficiency of this method, however, may be compromised if skipping RF encodings or $k_z$ planes to form the sampling pattern like that in blipped-CAIPI-enabled 2D SMS-EPI, because multiple $k_m$ locations cannot be covered during a single excitation.

An alternative strategy for SMSlab is to use a 3D k-space framework [25,27], which treats the concatenated thickness of all simultaneously excited slabs as a whole FOV for slice encoding. It is noteworthy that gaps exist between simultaneously excited slabs, which induce extra phase offsets under $k_z$ gradients, violating the Fourier encoding condition [25,27]. The inter-slab phase offsets can be removed to form the SMSlab 3D k-space, by using RF phase modulation along with $k_z$ encoding [25,27]. In this framework, CAIPI sampling can be formed by selecting complementary $k_y$ shots of different $k_z$ planes for interleaved EPI [25,27]. For single-shot EPI (ss-EPI) along $k_y$, however, blipped-CAIPI cannot be conducted to acquire multiple $k_z$ planes in one shot as in the single-slab 3D EPI case [30], because the inter-slab phase offsets vary with $k_z$ and only one $k_z$ can be phase-corrected in each shot through RF phase modulation.

In this study, we aim to apply blipped-CAIPI acquisitions for SMSlab (blipped-SMSlab) dMRI. A 4D k-space framework, which leverages the basic 4D k-space concept [31], is formulated for interpreting signal encoding. First, the SMSlab 4D k-space signal expression is formulated when using gradient encoding along both inter-slab and intra-slab dimensions. The interference between the two encodings is analyzed theoretically. Second, a blipped-SMSlab dMRI sequence is designed, with a multi-band 2D navigator for inter-kz-shot motion-induced phase correction in dMRI. Third, RF phase modulation and/or phase correction during reconstruction are implemented to decouple the inter-slab and intra-slab encodings that are otherwise entangled. In vivo experiments are performed to validate the proposed blipped-SMSlab method, and evaluate its performance in high-resolution dMRI compared with traditional 2D dMRI.

## 2 Theory

### 2.1 The 4D k-space signal expression of SMSlab

As shown in Figure 1, the SMSlab 4D k-space framework separates intra- and inter-slab encodings into two distinctive dimensions, namely $k_z$ and $k_m$. For intra-slab encoding, the resolution and FOV are defined as the slice thickness $\Delta z$ and slab thickness $FOV_z = N_z \cdot \Delta z$, respectively, where $N_z$ is the number of slices per slab (including over-sampled slices). For inter-slab encoding, i.e., along the multi-band dimension, the FOV is defined as $FOV_m = R_{mb} \cdot z_{\text{gap}}$, where $R_{mb}$ is the multi-band factor and $z_{\text{gap}}$ is the distance between the center of the simultaneously excited slabs. The 4D k-space and its image domain dimensions are represented by $k_x$-$k_y$-$k_m$-$k_z$ and x-y-m-z, respectively.

If the intra- and inter-slab encodings are independent of each other, the 4D k-space

signal can be expressed as

$$s(n_{km}, n_{kz}) = \sum_{n_z=0}^{N_z-1} \sum_{n_m=0}^{R_{mb}-1} \mu(n_m, n_z) e^{-i\frac{2\pi \cdot n_m \cdot n_{km}}{R_{mb}}} e^{-i\frac{2\pi \cdot n_z \cdot n_{kz}}{N_z}} \quad (1)$$

where $n_{km} \in [0, R_{mb} - 1]$ and $n_{kz} \in [-N_z/2, N_z/2 - 1]$ are the indices for $k_m$ and $k_z$ encodings, $n_m \in [0, R_{mb} - 1]$ is the index for simultaneously excited slabs, $n_z \in [0, N_z - 1]$ is the index for the slices per slab, $\mu(n_m, n_z)$ is the image intensity of the $n_z^{th}$ slice in the $n_m^{th}$ slab. The bottom slice in the bottom slab ($n_m = 0, n_z = 0$) is referred to as the "reference slice" below. The $k_x$ and $k_y$ dimensions are omitted for simplicity.

## 2.2 Influence of $k_m$ and $k_z$ gradients on the 4D k-space signal expression

In practice, however, the inter- and intra-slab encodings share the same physical z axis, which imposes extra phase interferences between them. Here, the case of $R_{mb} = 2$, i.e., two slabs are excited simultaneously (Slab 1 and 1' in Figure 2A), is used for illustration. For simplicity, the reference slice (Slice 1 in Figure 2A) is assumed to be at the isocenter (z=0).

The $k_z$ gradient produces a group of linear phases ranging $[0, 2\pi \cdot n_{kz}]$ across $FOV_z$ per slab along the z axis, which satisfies the intra-slab Fourier encoding condition. In the presence of the inter-slab center distance $z_{gap}$, however, an extra phase offset $\varphi_{gap}$ is introduced in the slab deviating from the isocenter (Slab 1'), which differs across different $k_z$ (Figure 2B):

$$\varphi_{gap} = 2\pi \cdot z_{gap}/FOV_z \cdot n_{kz} \cdot n_m \quad (2)$$

Similarly, the $k_m$ gradient (blipped-CAIPI gradient) imposes an extra linear ramp across different slices within each slab (Figure 2C):

$$\varphi_{ramp} = 2\pi \cdot \Delta z/FOV_m \cdot n_{km} \cdot n_z \quad (3)$$

According to the Fourier shift theorem, this phase ramp causes k-space location shift along $k_z$. For $R_{mb}$=2, the k-space samples with $n_{km} = 1$ slightly deviate from the Cartesian grid (Figure 2D), in a similar way as mentioned in a previous study [31].

Taking $\varphi_{gap}$ and $\varphi_{ramp}$ into consideration, Equation 1 should be modified as

$$s(n_{km}, n_{kz}) = \sum_{n_z=0}^{N_z-1} \sum_{n_m=0}^{R_{mb}-1} \mu(n_m, n_z) e^{-i\frac{2\pi \cdot n_m \cdot n_{km}}{R_{mb}}} e^{-i\frac{2\pi \cdot n_z \cdot n_{kz}}{N_z}} e^{-i \cdot \varphi_{\text{gap}}} e^{-i \cdot \varphi_{\text{ramp}}} \quad (4)$$

Taking $n_{kz} = 1$ and $n_{km} = 1$ for illustration, Figure 2E shows the compound phase at different slices per slab under $k_z$ and $k_m$ gradients. The $k_z$ gradient introduces an extra phase offset $\varphi_{\text{gap}}$ to Slab 1', while the $k_m$ gradient imposes an extra phase ramp within both slabs (the red line with an increased slope). To obtain signals that are not affected by $\varphi_{\text{gap}}$ and $\varphi_{\text{ramp}}$, the compound phase within each slab should follow the phase pattern in Figure 2G.

If the ratio between the inter-slab center distance $z_{\text{gap}}$ and $FOV_z$ is an integer, $\varphi_{\text{gap}}$ equals to an integer time of $2\pi$ according to Equation 2, which has no impact on the 4D k-space signal behavior. In SMSlab, slab over-sampling and overlap are necessary in most cases to reduce slab boundary artifacts even with correction [26,28], thus $z_{\text{gap}}/FOV_z$ is not an integer sometimes, which necessitates $\varphi_{\text{gap}}$ correction. Such an exception does not exist for $\varphi_{\text{ramp}}$, since $0 < \Delta z/FOV_m < 1$ in Equation 3 always holds in practice, thus additional correction is necessary.

When the reference slice is not at the isocenter (Supporting Information Figure S1), both the $k_z$ and $k_m$ gradients generate an extra phase with the off-isocenter distance $d_{\text{off\_iso}}$, namely off-isocenter phase error:

$$\varphi_{\text{off\_iso}} = \varphi_{\text{off\_iso,kz}} + \varphi_{\text{off\_iso,km}} = 2\pi \cdot \frac{d_{\text{off\_iso}}}{FOV_z} \cdot n_{kz} + 2\pi \cdot \frac{d_{\text{off\_iso}}}{FOV_m} \cdot n_{km} \quad (5)$$

The correction methods for $\varphi_{\text{gap}}$, $\varphi_{\text{ramp}}$ and $\varphi_{\text{off\_iso}}$ are provided in Sections 3.1.2/3.2.5, 3.2.2 and 3.2.1 in Methods, respectively.

## 3   Methods

### 3.1   Sequence Design

#### 3.1.1   The sequence of blipped-SMSlab dMRI

The blipped-SMSlab dMRI sequence is shown in Figure 3A. $k_z$ gradients are added before the EPI readout for intra-slab encoding. During the EPI readout of the image echo, which is single-shot for each $k_z$ plane, blipped-CAIPI gradients (i.e., $k_m$ gradients) are added between different EPI sub-echoes for inter-slab encoding, in a way similar to the original blipped-CAIPI implementation [10], except for that the $k_m$ encoding is designed to distribute the samples on integer $k_m$ coordinates when $R_{mb}$ is an even number. After the image-echo acquisition, reversed gradients are added to cancel the previous $k_z$ encodings. After the second refocusing pulse, a multi-band 2D navigator is acquired by treating each slab as a thick 2D slice to record the motion-induced phase variations of each $k_z$ shot.

The 4D k-space trajectory with blipped-CAIPI under-sampling is shown in Figure 3B. The traditional non-CAIPI under-sampling pattern (without $k_m$ gradients) is displayed in Figure 3C.

#### 3.1.2   RF phase modulation for $\varphi_{\text{gap}}$ correction

The $k_z$ gradient-induced inter-slab phase offset $\varphi_{\text{gap}}$ is analyzed in Section 2.2. It differs across different bands and different $k_z$, and can be calculated according to Equation 2. Two methods are proposed here for $\varphi_{\text{gap}}$ correction.

The first method directly removes $\varphi_{\text{gap}}$ during the acquisition. Since each $k_z$ is acquired in one shot, RF phase modulation is adopted to compensate for $\varphi_{\text{gap}}$ by introducing $-\varphi_{\text{gap}}$ to the multi-band RF pulses (Supporting Information, $\varphi_{\text{gap}}$ correction).

Without RF modulation, the second method removes $\varphi_{\text{gap}}$ during the image reconstruction, which is described in Section 3.2.5.

## 3.2 Reconstruction

The entire reconstruction pipeline is shown in Figure 4A and described below. The processing steps for b=0 s/mm² (linked by black arrows) and diffusion-weighted (DW) data (linked by blue arrows) are slightly different regarding $\varphi_{\text{ramp}}$ and inter-kz-shot phase correction. The reconstruction is implemented in MATLAB (Mathworks Inc., Natick, MA) on a PC computer with a 2.1 GHz Intel Xeon Gold CPU (22 cores) and 128 GB of RAM.

### 3.2.1 Off-isocenter phase error correction

The off-isocenter phase error $\varphi_{\text{off\_iso}}$, which is identical for all simultaneously excited slabs in one excitation, is removed during reconstruction according to Equation 5. Since the $k_z$ and $k_m$ gradients contribute to $\varphi_{\text{off\_iso}}$ differently, they should be handled separately. The $\varphi_{\text{off\_iso,km}}$ component varies across different $k_y$ lines due to blipped-CAIPI and is removed from the corresponding $k_y$ lines accordingly. While the $\varphi_{\text{off\_iso,kz}}$ component is consistent across different $k_y$ lines and is removed from all $k_y$ lines per $k_z$.

### 3.2.2 $\varphi_{\text{ramp}}$ correction

The $k_m$ gradient-induced intra-slab linear phase ramp $\varphi_{\text{ramp}}$ is analyzed in Section 2.2. Its correction is described here.

For b=0 s/mm² data, $\varphi_{\text{ramp}}$ is directly removed from the k-space signals after 1D inverse Fourier transform (iFT) along $k_z$.

For the DW data, since the motion-induced phase differs across different $k_z$ shots, inter-kz-shot phase variations should firstly be removed before 1D iFT along $k_z$. A three-step strategy is adopted for $\varphi_{\text{ramp}}$ correction of the DW data (Figure 4B):

(1) remove inter-kz-shot phase (see Section 3.2.4 for details), which requires a preliminary image reconstruction step for each $k_z$ plane;

(2) remove $\varphi_{\text{ramp}}$ after 1D iFT along $k_z$;

(3) add the inter-kz-shot phase variation back and return to the under-sampled k-space, then replace the original samples contaminated by $\varphi_{\text{ramp}}$ with the corrected samples (bottom panel of Figure 4B).

### 3.2.3 Nyquist ghost correction

The Nyquist ghost is corrected using the acquisition-free singular value decomposition (SVD) based algorithm[32]. The b=0 s/mm² images with $k_z = 0$ are used to determine the optimal constant and linear phases, which are then applied to all $k_z$ planes of the b=0 s/mm² and diffusion scans.

Note that when using the SVD based method, the ghost correction of the b=0 s/mm² data should be performed after removing $\varphi_{\text{ramp}}$ (Figure 4A), since $\varphi_{\text{ramp}}$ also differs across even and odd EPI echoes, which affects the phase estimation for ghost correction.

For the DW data, ghost correction is carried out before $\varphi_{\text{ramp}}$ correction, because there is a preliminary image reconstruction during $\varphi_{\text{ramp}}$ correction (Figure 4B, Step 1).

### 3.2.4 2D-GRAPPA and inter-kz-shot phase correction

For b=0 and DW images, a two-step 2D-GRAPPA-operater[33] is adopted to reconstruct each $k_z$ plane. The interpolation kernel is shown in Supporting Information Figure S2B. The 2D-GRAPPA calibration data are obtained from a fully-sampled 2D interleaved EPI scan with b=0 s/mm², treating the simultaneously excited slabs as thick slices and performing 3D FT (Supporting Information Figure S2A).

For the b=0 s/mm² scan, the final images are obtained using 4D iFT after 2D-GRAPPA.

For the diffusion scans, inter-kz-shot motion-induced phase correction is performed before iFT along $k_z$ using the navigator information. The correction procedures are described below and shown in Supporting Information Figure S3, with interim magnitude and phase images shown in Supporting Information Figure S4. For each $k_z$ shot, the corresponding multi-band 2D navigator images are firstly reconstructed and coil combined. Then, the navigator phase maps are extracted and convolutionally filtered using a Gaussian kernel with a width of 10 pixels and the full width at half maximum (FWHM) of 4 pixels. The phase maps are then applied to the reconstructed images of each $k_z$ plane. The final diffusion images are obtained through 1D iFT along $k_z$.

When using partial Fourier acquisition along $k_y$, the POCS-based method is used for partial Fourier reconstruction [34] after 2D-GRAPPA and inter-kz-shot phase correction.

### 3.2.5 The second method for $\varphi_{\text{gap}}$ correction

Without RF phase modulation, the $k_z$ gradient-induced inter-slab phase offset $\varphi_{\text{gap}}$ can also be corrected during reconstruction. Since $\varphi_{\text{gap}}$ varies among different bands and $k_z$ encodings, it can be treated in a similar way compared with the inter-kz-shot motion-induced phase (see implementation details in Supporting Information, $\varphi_{\text{gap}}$ correction).

### 3.2.6 Other artifacts correction

After the aforementioned reconstruction procedures, CPEN [28] is performed to correct for slab boundary artifacts, using slab profiles estimated from the calibration scan in Section 3.3.5. Then, eddy-current effects are corrected using the *eddy* function [35] from the FSL software package (v6.0.3) [36]. B0 inhomogeneity-induced distortions are corrected using the B0 field map estimated from the FSL's *topup* function [37] and an optimized Jacobian modulation matrix [38].

### 3.3 Data Acquisition

Whole-brain data were acquired on a Philips 3.0T Ingenia CX MR scanner (Philips Healthcare, Best, The Netherlands) equipped with a 32-channel head coil. The maximum gradient strength and slew rate are 80 mT/m and 100 T/m/s, respectively. The study was approved by the local Institutional Review Board and written informed consents were obtained from three healthy volunteers. The root-flipped RF pulse design [39-41] was used to shorten TE.

Four experiments were carried out to test the efficacy of blipped-SMSlab, and compare its performance with traditional 2D methods for dMRI. For blipped-SMSlab, 10 to 14 slabs were acquired with different resolutions to cover the whole brain. For each slab, 8 to 12 $k_z$ planes were acquired, which resulted in 8 to 12 slices, including one over-sampled slice on each side of a slab. Therefore, adjacent slabs overlapped by 2 slices. Each $k_z$ plane was acquired using ss-EPI with $R_{mb} = 2$ and $R_y = 2$, which together formed the blipped-CAIPI sampling trajectories as shown in Figure 3B. The 2D navigator with $R_{mb} = 2$ had a matrix size 84×29, without $k_y$ acceleration. Some common imaging parameters were: FOV$_{xy}$ = 220×220 mm$^2$, excitation/refocusing pulse flip angle=90°/180°. Partial Fourier was applied along $k_y$ but not $k_z$. Other detailed parameters are described below and listed in Table 1.

### 3.3.1 Experiment 1

First, the efficacy of RF phase modulation for $\varphi_{\text{gap}}$ correction was evaluated, with one volunteer undergoing scans 1 and 2 (Table 1), without and with RF phase modulation, respectively. The two scans differed in the ratio $z_{\text{gap}}/FOV_z$, which is a non-integer number in scan 1 and an integer number in scan 2. When RF phase modulation was not used, the feasibility of the second $\varphi_{\text{gap}}$ correction method performed in reconstruction was tested. The efficacy of the $\varphi_{\text{ramp}}$ correction method was also evaluated.

### 3.3.2 Experiment 2

The performance of the blipped-CAIPI sampling pattern and the non-CAIPI sampling pattern was compared, with the same volunteer undergoing scans 2 and 3 (Table 1), respectively.

### 3.3.3 Experiment 3

Blipped-SMSlab, blipped-CAIPI-enabled 2D SMS-EPI and single-band 2D EPI (scans 4~9, Table 1) were compared, using single-shot acquisition in the $k_x$-$k_y$ plane with $R_y = 2$. DTI data were acquired from two volunteers at 1.3 mm (the 1$^{st}$ volunteer) or 1.0 mm (the 2$^{nd}$ volunteer) isotropic resolution, with b=1000 and b=2000 s/mm$^2$ and 6 diffusion directions for each b-value. To match the acquisition time, a single average was acquired for SMSlab and single-band 2D EPI, and 2 averages were acquired for 2D SMS-EPI. Another b=0 s/mm$^2$ scan was conducted for each acquisition with reversed phase encoding gradient for B0 inhomogeneity-induced distortion correction.

### 3.3.4 Experiment 4

Blipped-SMSlab and 2D SMS-EPI dMRI were compared at 1.0 mm isotropic resolution with b=1000 s/mm$^2$ (scans 10~12, Table 1). To match the acquisition time, blipped-SMSlab data were acquired along 32 diffusion directions with 1 average, while 2D SMS-EPI data were acquired along 32 diffusion directions with 2 averages, and 64 diffusion directions with 1 average, from the 3$^{rd}$ volunteer. Another b=0 s/mm$^2$ scan was conducted for each acquisition with reversed phase encoding gradient for B0 inhomogeneity-induced distortion correction.

### 3.3.5 Calibration scans

A 2-shot 2D interleaved EPI sequence was used to obtain the 2D-GRAPPA calibration data (b=0 s/mm$^2$) as well as the coil sensitivity map for coil combination (Supporting Information Figure S2A). A slab profile calibration scan was conducted using the same sequence as the SMSlab b=0 s/mm$^2$ scan, but with a 2-fold over-sampling along the slice direction for each slab. The slab profiles were calculated for CPEN correction as described before [28]. Detailed imaging parameters of the calibration scans are listed in Supporting Information Table S1.

## 3.4 Data Analysis

SNR and g-factor maps were calculated using the pseudo-multiple replica method [42] with 128 repetitions. Fractional anisotropy (FA) and mean diffusivity (MD) were calculated using the FSL's *dtifit* function [36]. Fiber orientation distribution function (ODF) was estimated using the combined general orientation transform with constrained optimization (coGOT) method [43].

## 4 Results

Figure 5 shows the blipped-SMSlab images from scan 1 in Experiment 1, without and with the correction for the $k_z$ gradient-induced inter-slab phase offset $\varphi_{\text{gap}}$, respectively. In this scan, the ratio between the inter-slab center distance $z_{\text{gap}}$ and $FOV_z$ is not an integer. Image shift occurs in Slab 1' without $\varphi_{\text{gap}}$ correction (Figure 5A), and the shift distance is equal to the result of $z_{\text{gap}}$ modulo $FOV_z$. This problem can be addressed with either RF phase modulation or correction during reconstruction alone. Figure 5B shows an example result using RF phase modulation, while the second correction method yields the same $\varphi_{\text{gap}}$ correction result (Supporting Information Figure S5A). When $z_{\text{gap}}/FOV_z$ is an integer, the images can be accurately reconstructed even without $\varphi_{\text{gap}}$ correction (Supporting Information Figure S5B).

Figure 6 shows the blipped-SMSlab images with and without correction of the $k_m$ gradient-induced intra-slab linear phase ramp $\varphi_{\text{ramp}}$. Without $\varphi_{\text{ramp}}$ correction, aliasing artifacts appear, along both the multi-band direction (e.g., areas highlighted by red circles in Figure 6C) and the phase encoding direction within each slice (areas highlighted by yellow arrows). These artifacts are substantially ghost artifacts that arises from: (1) sampling differences between different $k_m$ encodings (i.e., slight shift exists between different $k_m$ locations, Figure 2D); (2) sampling differences between different $k_y$, which are with different $k_m$ encodings due to the blipped-CAIPI trajectory. Specifically, in the $R_{mb} = 2$ case, there is no extra phase caused by the $k_m = 0$ encoding, whereas there is extra intra-slab phase ramp caused by the $k_m \neq 0$ encoding gradient along z. The $\varphi_{\text{ramp}}$-induced ghosting level grows along z within each slab, up to 14.8%, because the inter-$k_y$ phase difference increases for slices far from the bottom slice within each slab.

The reconstructed diffusion images with and without inter-kz-shot motion-induced phase correction are shown in Figure 7. Without correction, the diffusion images are corrupted. They are well recovered after correction using the navigator information. The diffusion images obtained before and after slab boundary artifacts correction as well as distortion correction are shown in Supporting Information Figure S6.

The g-factor for blipped-CAIPI and non-CAIPI acquisitions are assessed using data from Experiment 2. Quantitatively, the whole-brain averaged g-factors are 1.44 ± 0.26 (mean ± standard deviation, across all voxels in the brain region of a single subject) and 1.63 ± 0.44 for the blipped-CAIPI and non-CAIPI SMSlab acquisitions, respectively, indicating a 11.7% reduction for blipped-SMSlab imaging. The whole-brain averaged SNRs are 16.5 ± 9.5 and 14.8 ± 8.6 for the blipped-CAIPI and non-CAIPI SMSlab acquisitions, indicating a 12% increase for blipped-SMSlab imaging. The g-factor and SNR maps of two example slices are provided in Supporting Information Figure S7.

Figure 8 compares dMRI images acquired using blipped-SMSlab, 2D SMS-EPI and 2D EPI in Experiment 3. The single-direction b=1000 s/mm$^2$ and b=2000 s/mm$^2$ images, FA and MD maps, at 1.3 mm and 1.0 mm isotropic resolutions are shown. For each resolution, the acquisition time keeps consistent across all three scans. Quantitatively, for the 1.3 mm isotropic single-direction DW images, the whole-brain averaged SNRs of blipped-SMSlab (before CPEN), 2D SMS-EPI and 2D EPI are 5.81 ± 2.41 (mean ± standard deviation, across all voxels in the brain region of a single subject), 5.48 ± 2.37 and 4.23 ± 1.74 for b=1000 s/mm$^2$, 3.32 ± 1.63, 3.18 ± 1.56 and 2.43 ± 1.14 for b=2000 s/mm$^2$. For the 1.0 mm isotopic images, the SNRs of blipped-SMSlab, 2D SMS-EPI and 2D EPI are 3.05 ± 1.41, 2.48 ± 1.24 and 2.09 ± 0.94 for b=1000 s/mm$^2$, 1.75 ± 0.93, 1.56 ± 0.84 and 1.29 ± 0.65 for b=2000 s/mm$^2$, respectively. Note that the SNR map of the two-average 2D SMS-EPI was calculated first from a single average and then scaled with $\sqrt{2}$. The result shows 4.4~6.0% SNR improvement of blipped-SMSlab over 2D SMS-EPI for 1.3 mm isotropic imaging, and 12.2~23.0% for 1.0 mm isotropic imaging, which indicates a growing SNR advantage of blipped-SMSlab for higher resolution imaging. This trend is consistent with the observation of a previous work [9].

Figure 9 further compares the color-coded FA (cFA) maps and ODF from blipped-SMSlab and two kinds of 2D SMS-EPI acquisitions at 1.0 mm isotropic resolution in Experiment 4, using matched acquisition time. Due to higher SNR efficiency, the blipped-SMSlab shows higher SNR of the cFA maps and better organized ODFs, compared with both the 2D SMS-EPI acquisitions. The difference is more prominent in areas with much lower SNR (indicated by read rectangles).

In terms of off-line reconstruction time per diffusion direction, it takes about 2 minutes for 2D EPI and 2D SMS-EPI to reconstruct a whole brain volume with 84 slices at 1.5 mm isotropic resolution. The non-CAIPI SMSlab takes about 4 minutes, which includes additional time for navigator processing and inter-kz-shot phase correction compared with the 2D methods. The blipped-SMSlab takes about 5 minutes when using RF modulation to remove $\varphi_{\text{gap}}$, 5.4 minutes when without RF modulation and removing $\varphi_{\text{gap}}$ during reconstruction. The blipped-SMSlab reconstruction takes a bit longer than non-CAIPI SMSlab mainly due to the $\varphi_{\text{ramp}}$ correction. Note that both the blipped-CAIPI and non-CAIPI SMSlab needs additional 1~3 minutes per diffusion direction for slab boundary artifact correction using CPEN [28]. The computations are not currently optimized to save memory and time, which can be much faster using coil compression [44] to reduce data size and/or using parallel computing.

## 5 Discussion

In this study, the SNR-efficient simultaneous multi-slab imaging with blipped-CAIPI (blipped-SMSlab) in a 4D k-space framework was proposed and evaluated for dMRI. The 4D k-space signal formulation for SMSlab was established. The extra phases induced because inter-slab and intra-slab encodings share the same physical z axis were analyzed. The extra phases were corrected by RF modulation and correction during reconstruction, or correction during reconstruction alone, thus decoupling inter-slab and intra-slab encodings and enabling the blipped-CAIPI acquisition. For dMRI acquisitions, the inter-kz-shot phase variations were removed using the phase information from the multi-band 2D navigator. In vivo experiments validated the efficacy of the proposed blipped-SMSlab method, demonstrated the advantage of the blipped-CAIPI sampling pattern over the non-CAIPI sampling pattern regarding g-factor-related SNR penalty, as well as the SNR advantage of blipped-SMSlab over 2D dMRI.

In the 4D k-space framework, if without correcting the $k_z$ gradient-induced inter-slab phase offset $\varphi_{\text{gap}}$, slice shift occurs along the z direction within Slab 1'. The shifted slices have no interaction with Slab 1 (Figure 5A), which allows $\varphi_{\text{gap}}$ removal using either RF phase modulation or correction during reconstruction. This phenomenon is different from that in the SMSlab 3D k-space framework [25,27] (Supporting Information Figure S8D), in which several slices in Slab 1' shift to Slab 1, and they alias together with the slices in Slab 1 [27]. Consequently, RF phase modulation is required for the SMSlab 3D k-space framework using k-space-based reconstruction, since either removing the inter-slab phase offset during reconstruction or shifting aliased slices back after reconstruction is difficult to realize.

When correcting slab boundary artifacts, it should be noticed that the artifact pattern of blipped-SMSlab in the 4D k-space framework is different from that of the SMSlab 3D k-space framework, regarding different boundary slice aliasing patterns due to different definitions of FOV. The slab boundary aliasing pattern of blipped-SMSlab is the same as that in single-band multi-slab imaging, in which only intra-slab aliasings exist at the edge slices (Supporting Information Figure S9), and the single-band multi-slab CPEN correction model should be selected accordingly [28].

Compared with SMSlab using non-CAIPI sampling [9,24], blipped-SMSlab can reduce the g-factor penalty, by introducing $FOV_y$ shift between simultaneously excited slabs to improve the utility of coil sensitivity variation. In the previous 3D k-space framework [25,27], when using ss-EPI along $k_y$ and skipping some of the $k_z$ encodings for acceleration, the parallel imaging reconstruction performance is similar to the non-CAIPI sampling pattern in the 4D k-space framework, which cannot fully utilize the coil sensitivity information.

The whole-brain volume acquisition time of blipped-SMSlab dMRI along each diffusion direction is about 20 s (TR=1.6~2.1s), which is about twice that of 2D SMS-EPI dMRI. The longer volume acquisition time of blipped-SMSlab is mainly due to additional navigator acquisitions, which requires about 50% additional time. Another factor is the minor slice oversampling adopted for high slab boundary artifact correction performance [26,28], which takes about 20~30% more acquisition time. Self-navigation for SMSlab dMRI has been proposed to avoid navigator acquisition [9], but was not adopted in this study since the data were noisy especially for large $k_z$ encoding gradients. To match the total acquisition time for fair comparison, SMSlab was acquired with 1 average, while 2D SMS-EPI was acquired with two averages or double diffusion directions. The one-average blipped-SMSlab dMRI shows higher SNR than the two-average or double-diffusion-direction 2D SMS-EPI dMRI due to higher SNR efficiency (Figures 8 and 9), even after denoising (Supporting Information Figure S10).

With advanced receive coil arrays, the SNR efficiency of 2D SMS-EPI dMRI may be further improved with higher multi-band factors to shorten TR to the optimal range, with a side benefit of reduced volume acquisition time. However, this manner is at the cost of amplified g-factor penalties. In this scenario, even using multiple averages to match the total acquisition time, the SNR of 2D SMS-EPI is still theoretically compromised compared with the TR-optimized blipped-SMSlab using $R_{mb} = 2$. The volume acquisition time of blipped-SMSlab can also be reduced using thinner slabs, together with slightly higher multi-band factors (if a multi-channel receive array supports) to maintain the optimal TR, yet the g-factor issue should also be carefully balanced.

The 2D and 3D methods are distinctively affected by the non-ideal slice or slab profile of the RF pulses. In 2D imaging, image blurring is directly induced by the non-ideal slice profile. In 3D multi-slab and SMSlab, image blurring is indirectly induced during slab boundary artifact correction, which is related to the regularization term or loss function used to suppress slab boundary artifacts, as evaluated in a previous study [28]. For the 3D Fourier-based imaging, the image reconstruction using the truncated Fourier series has the underlying SINC-like point spread function (PSF) [45] along all encoding directions, as shown in Supporting Information Figure S11. The PSF of Fourier reconstruction and subsequent processing procedures, such as CPEN correction and potentially imperfect inter-kz-shot phase correction, may introduce blurring along the slice direction. The blurring effect of the 2D and 3D data (after CPEN) in this study was preliminarily evaluated by estimating the FWHM of the actual PSF using the FSL's smoothest function [36], which estimates the FWHM of the Gaussian kernel needed to smooth white noise to have the same "smoothness" as the data. The estimated FWHM of the 3D data is about 20% larger than the 2D data. The blurring effect and the truncation artifacts along the slice direction [46] are potential limitations of the 3D method, which will be further investigated in future work.

In multi-slab and SMSlab imaging, bulk motion between different $k_z$ shots can cause substantial signal variations [15]. Moreover, with short TRs, bulk motion between the excitation of different slabs may also cause motion-related spin-history effects. This phenomenon has been investigated for 2D dMRI [47,48], in which motion-related spin-history effects cause signal modulation of some slices with TR<3s. Bulk motion is not considered in this study with cooperative volunteers, which will be further investigated and corrected in future work.

# 6 Conclusion

In this study, an SNR-efficient simultaneous multi-slab method with blipped-CAIPI (blipped-SMSlab) in a 4D k-space framework is developed for 3D high-resolution whole-brain diffusion imaging at 3T. The phase problem from the inter-slab and intra-slab gradient encodings along the same physical z axis are addressed, thus enabling blipped-CAIPI acquisitions. The blipped-CAIPI sampling has been demonstrated to be superior to non-CAIPI sampling regarding g-factor and SNR. High-resolution (1.0 mm isotropic) 3D diffusion images have been obtained using the proposed method, which outperforms traditional 2D dMRI methods regarding SNR for matched resolution and acquisition time and thus capable of high-quality, high-resolution fiber orientation detection.

**Acknowledgements**

The authors would like to thank the reviewers for insightful comments and suggestions concerning the evaluation of the proposed method, Dr. Zhi-Pei Liang at University of Illinois at Urbana-Champaign for helpful discussions about the SNR and image blurring topics, Dr. Ying Chen for helpful discussions about the 4D k-space theory, Dr. Erpeng Dai at Stanford University for helpful discussions about g-factor calculation, Dr. Qiyuan Tian at Massachusetts General Hospital for proofreading the manuscript, Dr. Fuyixue Wang at Massachusetts General Hospital for helpful discussions about the point spread function, and Mr. Xin Shao at Tsinghua University for assistance with data acquisition. This work was supported by Beijing Municipal Natural Science Foundation (L192006) and the National Natural Science Foundation of China (61971258).

**Figures:**

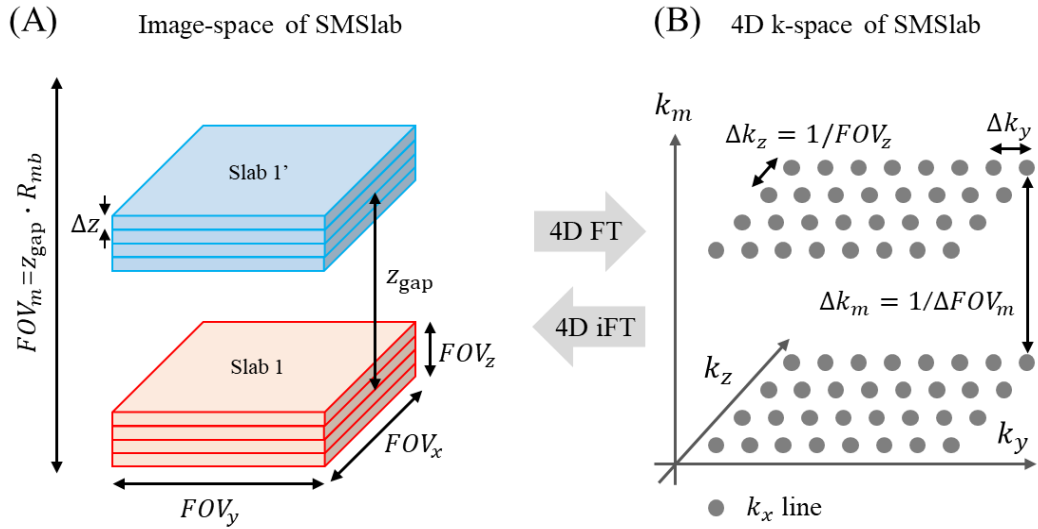

Figure 1. The 4D k-space framework of SMSlab imaging ($R_{mb}$=2 for illustration). (A) SMSlab excitation and the thin slices to be reconstructed in the image domain. $z_{gap}$ is the center distance between the simultaneously excited slabs (Slab 1 and 1'). (B) The 4D k-space, where the $k_x$ dimension is omitted and each gray point represents a whole $k_x$ line. 4D FT: 4D Fourier transform.

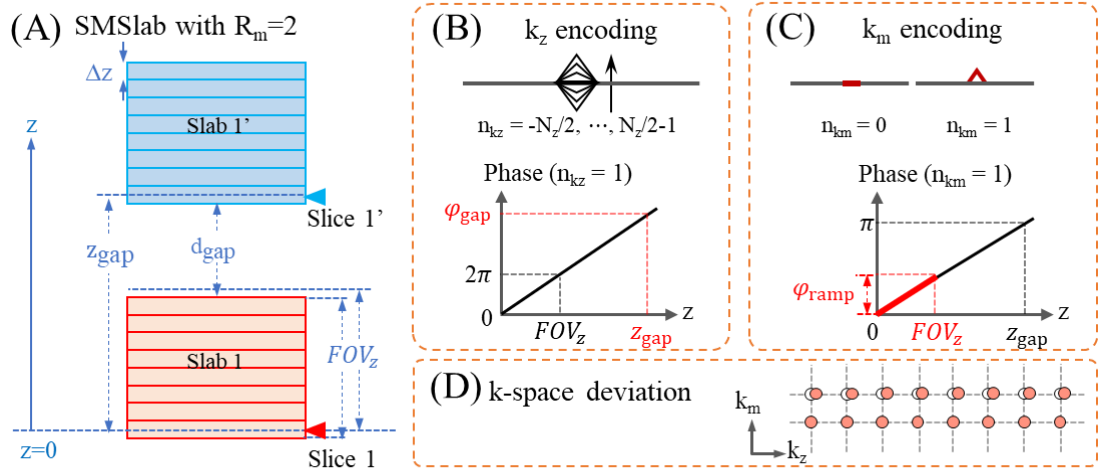
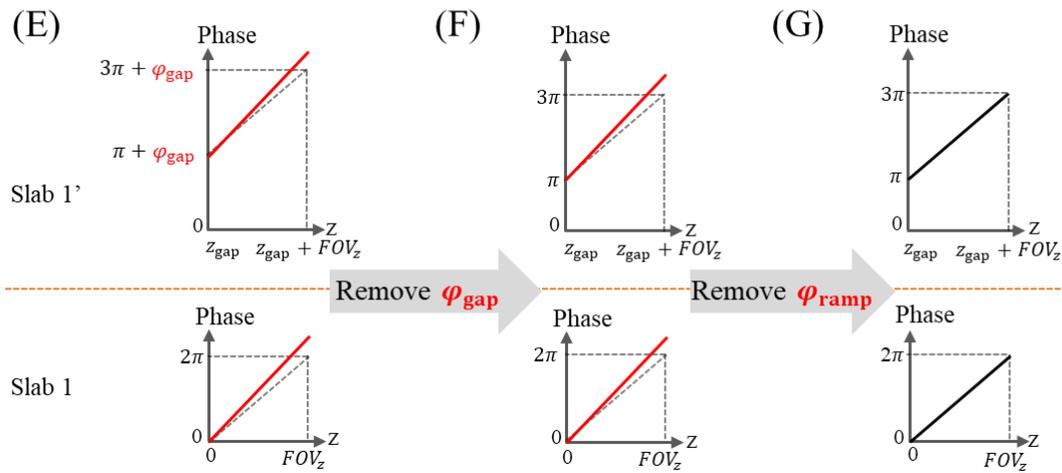

Figure 2. Illustration of extra phase interferences caused by the $k_m$ and $k_z$ gradients. (A) SMSlab excitation ($R_{mb}$=2 for illustration) when the reference slice (Slice 1) is at the isocenter (z=0). (B) The phase along z axis under $k_z$ encoding ($n_{kz}$=1 for illustration). (C) The phase along z axis under $k_m$ encoding ($n_{km}$=1 for illustration). (D) Illustration of k-space location deviation due to the $k_m$ gradient applied. (E-G) Compound phase under $k_z$ & $k_m$ encodings within each slab along the z axis without any correction, with $\varphi_{gap}$ correction and further with $\varphi_{ramp}$ correction, respectively.

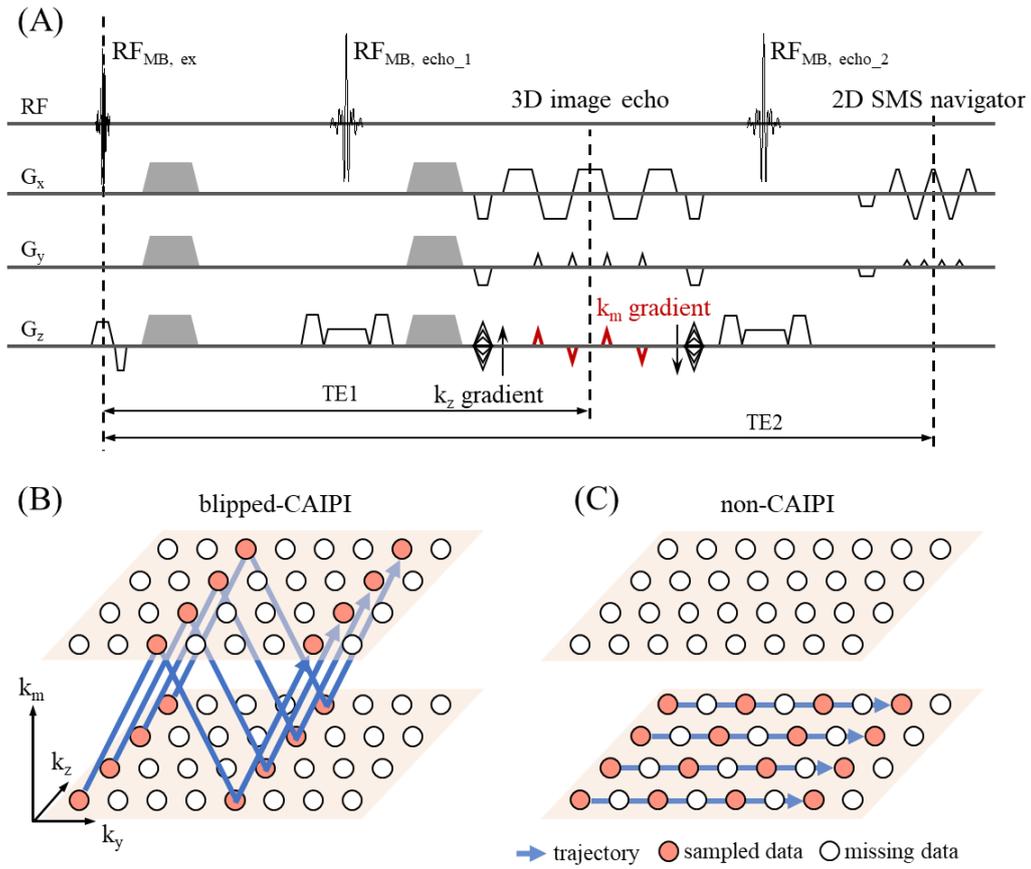

Figure 3. (A) The sequence diagram of the blipped-SMSlab dMRI method. (B) Sampling trajectory with blipped-CAIPI in the 4D k-space ($R_{mb}$=2, $R_y$=2). (C) Sampling trajectory without blipped-CAIPI in the 4D k-space ($R_{mb}$=2, $R_y$=2).

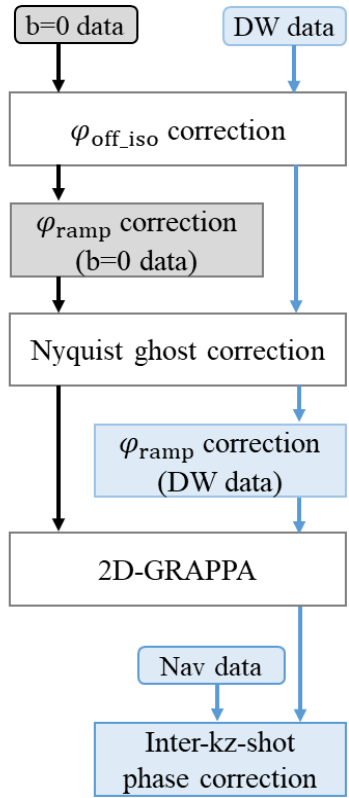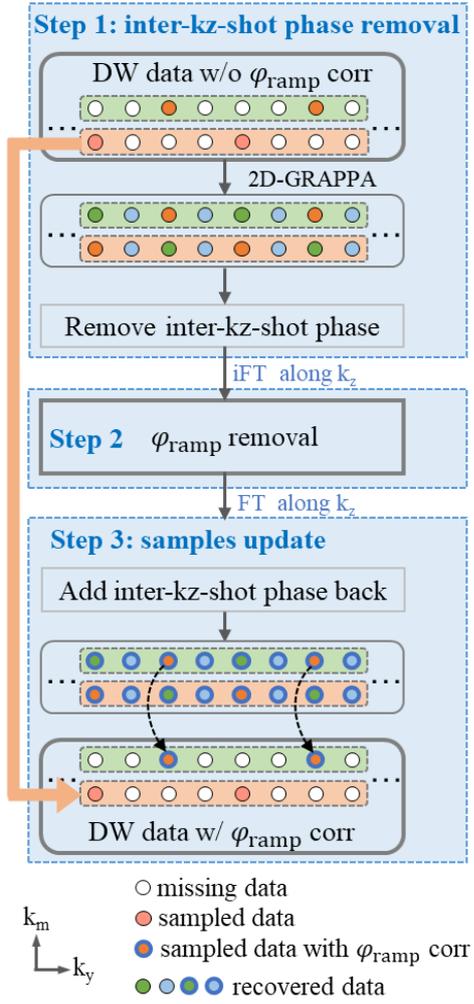

Figure 4. (A) Reconstruction pipeline of blipped-SMSlab dMRI. (B) Correction of the $k_m$ gradient-induced intra-slab linear phase ramp $\varphi_{ramp}$ for diffusion data.

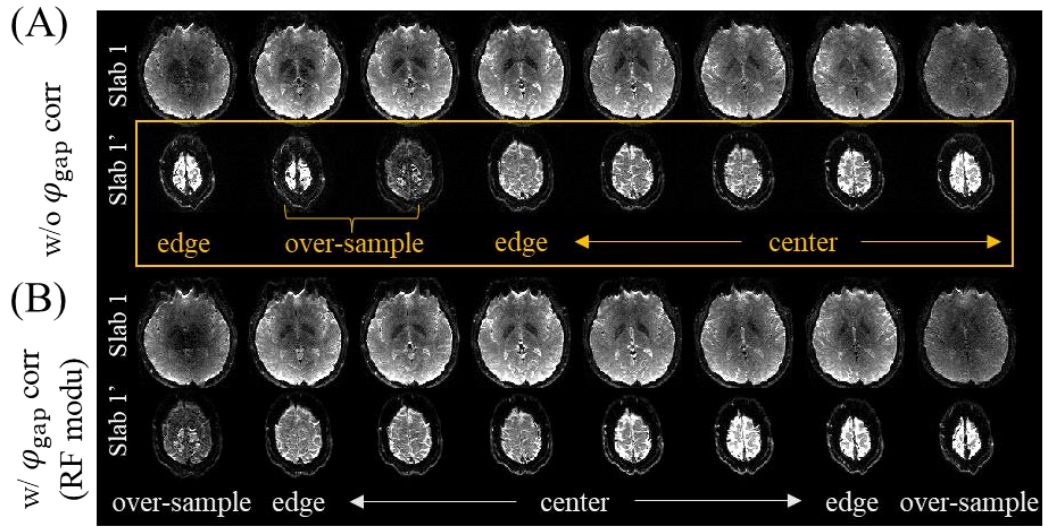

Figure 5. The blipped-SMSlab b=0 s/mm² images without and with $\varphi_{gap}$ correction, respectively. In this case (scan 1 in Table 1), the ratio between the center distance of the simultaneously excited slabs ($z_{gap}$) and $FOV_z$ is not an integer, and $z_{gap}$ modulo $FOV_z$ equals to 2. When without correction (A), image shift occurs in Slab 1', which is limited to Slab 1' and does not contaminate Slab 1. This problem can be addressed with either RF modulation (B) or correction during reconstruction alone (Supporting Information Figure S5A).

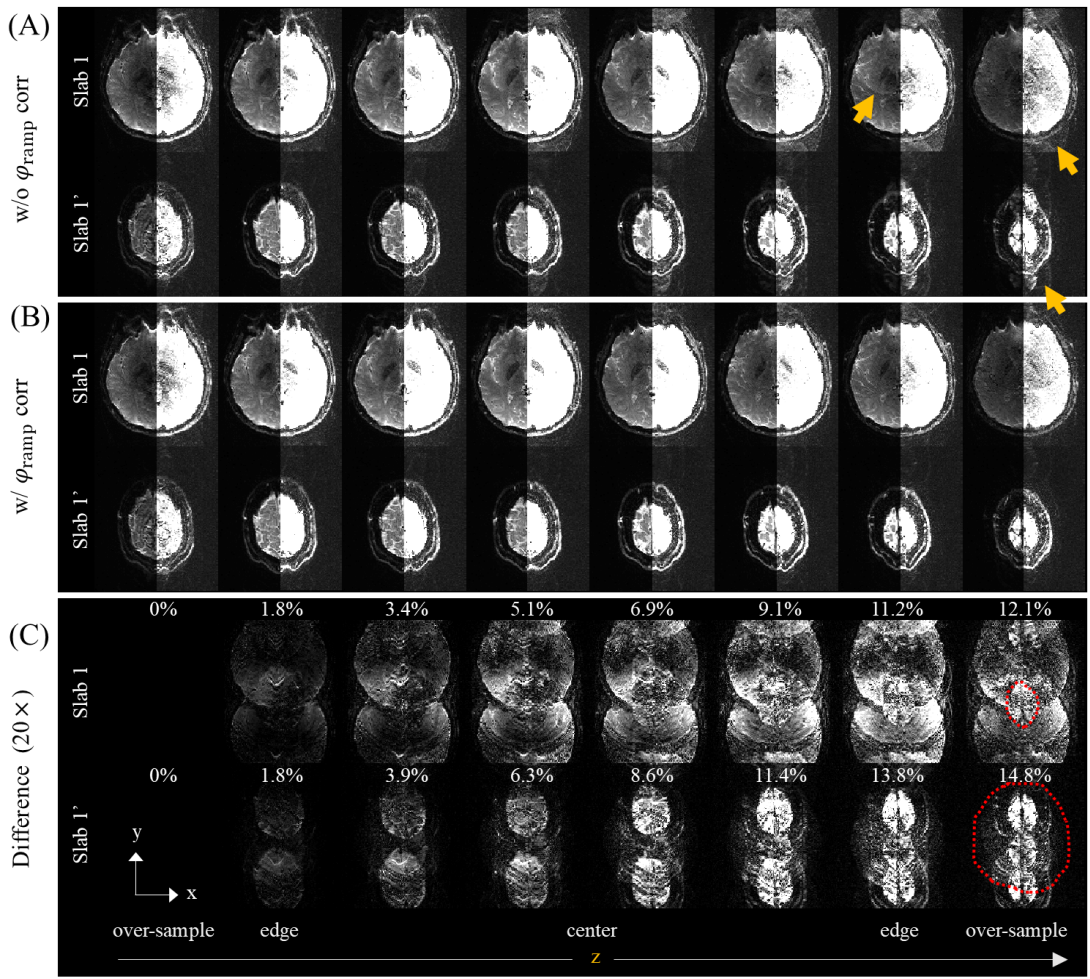

Figure 6. The blipped-SMSlab images without (A) and with $\varphi_{\text{ramp}}$ correction (B). (C) shows the 20× difference map between panels A and B. In panels A and B, the right-half of each slice shows the lowest 20% of the signals. If without correction, $\varphi_{\text{ramp}}$ causes inter-slab ghost artifacts, which can be more clearly observed in the example areas highlighted by the dotted red circles in the difference map (C). With blipped-CAIPI sampling, $\varphi_{\text{ramp}}$ also causes intra-slice ghost artifacts along the phase encoding direction, as shown in the areas highlighted by the yellow arrows. Within each slab, the ghosting level grows as the phase $\varphi_{\text{ramp}}$ ramps up along the z direction. The percentage of ghosting is calculated through dividing the mean intensity of the difference map by the mean intensity of the $\varphi_{\text{ramp}}$ corrected images. The left-most slice in Slab 1 corresponds to the reference slice (Slice 1) in Figure 2A. These ghost artifacts can be removed with proper $\varphi_{\text{ramp}}$ correction (B).

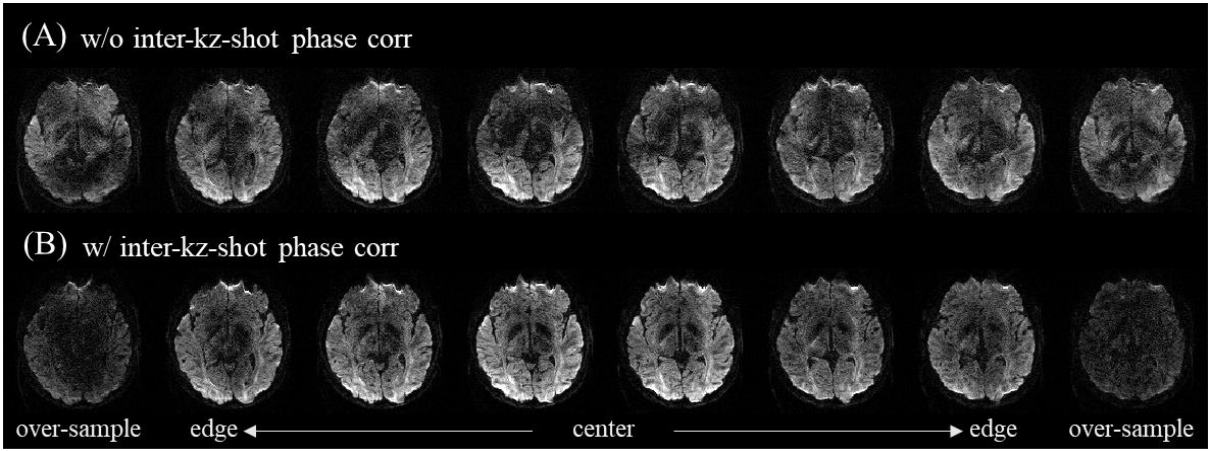

Figure 7. The single-direction blipped-SMSlab diffusion images without (A) and with inter-kz-shot phase correction (B). All slices in one slab, including the oversampled slices, are shown before slab boundary artifact correction. Without inter-kz-shot phase correction, artifacts from physiological motion corrupts the diffusion images (A). With multi-band 2D navigator-based phase correction, the images can be correctly reconstructed (B).

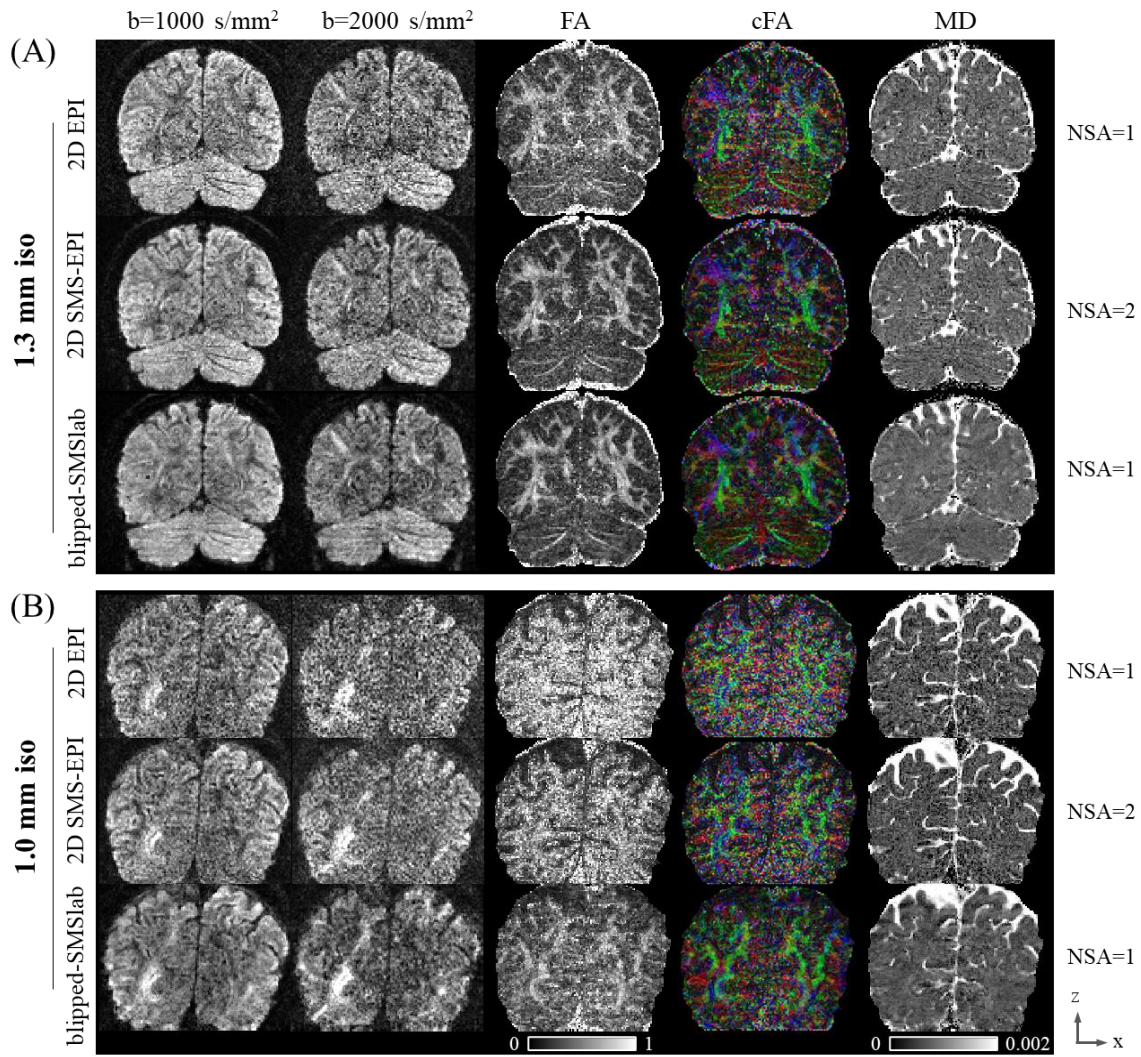

Figure 8. Comparison between blipped-SMSlab, 2D SMS-EPI with blipped-CAIPI and 2D EPI, for single-direction b=1000 s/mm² and b=2000 s/mm² images, FA and MD maps. (A) The imaging result of the first volunteer with 1.3 mm isotropic resolution. (B) The imaging result of the second volunteer with 1.0 mm isotropic resolution. For each resolution scenario, the acquisition time keeps consistent across the three acquisitions (3:54 and 4:10 for 1.3 mm and 1.0 mm resolutions, respectively). The images shown here are after slab boundary artifact correction and distortion correction.

(A) 2D SMS-EPI, $N_{dir}$=32, NSA=2

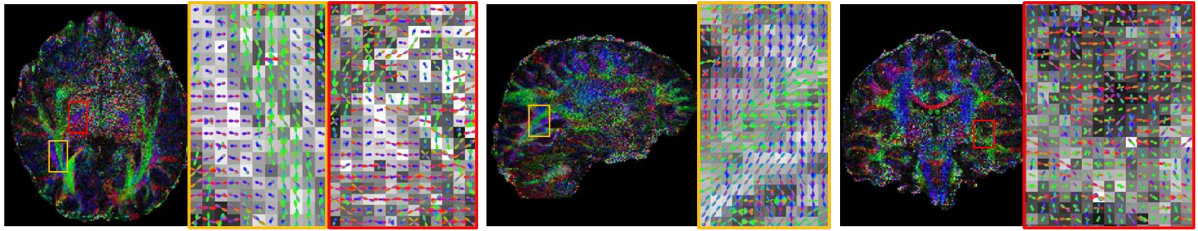

(B) 2D SMS-EPI, $N_{dir}$=64, NSA=1

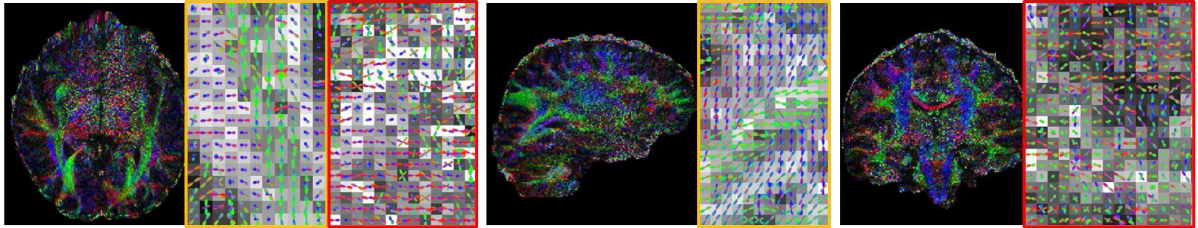

(C) blipped-SMSlab, $N_{dir}$=32, NSA=1

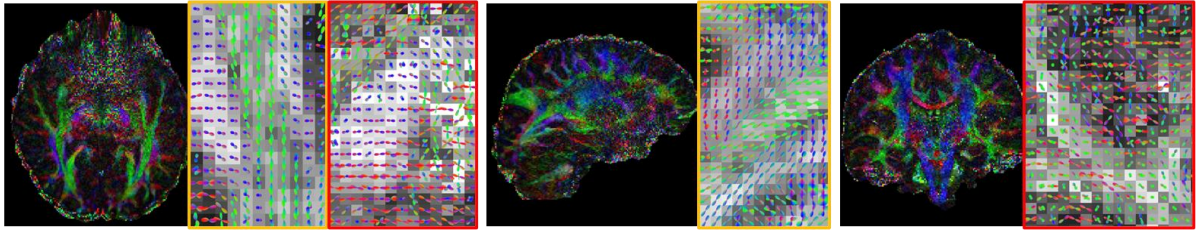

Figure 9. Comparison of color-coded FA maps and ODF between blipped-SMSlab and 2D SMS-EPI using blipped-CAIPI, with 1.0 mm isotropic resolution. (A) 2D SMS-EPI dMRI with 32 diffusion directions and 2 averages. (B) 2D SMS-EPI dMRI with 64 diffusion directions and 1 average. (C) The blipped-SMSlab dMRI with 32 diffusion directions and 1 average. The acquisition time keeps consistent across the three acquisitions (~14 min for each). $N_{dir}$: number of diffusion directions.

## Tables

Table 1. Acquisition parameters of SMSlab dMRI and 2D dMRI.

| Exp. # | Sub. # | Scan # (type) | Res. (mm³) | $R_{mb}\times$ Slabs | $N_z$ ($N_{target}\times OS$)[a] | $\dfrac{z_{gap}}{FOV_z}$ | b value (s/mm²) ($N_{dir}$) | PF | TE1/ TE2 (ms) | TR (s) | Time per volume (s)[c] | NSA | Total scan time (min:s)[d] |
|---|---|---|---|---|---|---|---|---|---|---|---|---|---|
| 1 | 1 | 1 (SMSlab, blipped-CAIPI) | 1.5³ | 2×7 | 8 (6×1.33) | 5.25 | 0 (1), 1k (6) | 0.7 | 77/180 | 1.8 | 14.4 | 1 | 1:41 |
|   |   | 2 (SMSlab, blipped-CAIPI) | 1.5³ | 2×5 | 10 (8×1.25) | 4 | 0 (1), 1k (6) | 0.7 | 77/180 | 1.8 | 18 | 1 | 2:06 |
| 2 | 1 | 3 (SMSlab, non-CAIPI) | 1.5³ | 2×5 | 10 (8×1.25) | 4 | 0 (1), 1k (6) | 0.7 | 77/180 | 1.8 | 18 | 1 | 2:06 |
| 3 | 1 | 4 (SMSlab, blipped-CAIPI) | 1.3³ | 2×7 | 10 (8×1.25) | 5.6 | 0 (1, +1 AP[b]), 1k (6), 2k (6) | 0.65 | 85/200 | 1.8 | 18 | 1 | 3:54 |
|   |   | 5 (2D SMS, blipped-CAIPI) | 1.3³ | 2×56 | 1 |   | 0 (1, +1 AP), 1k (6), 2k (6) | 0.65 | 85/-- | 9 | 9 | 2 | 3:54 |
|   |   | 6 (2D) | 1.3³ | 1×112 | 1 |   | 0 (1, +1 AP), 1k (6), 2k (6) | 0.65 | 85/-- | 18 | 18 | 1 | 3:54 |
|   | 2 | 7 (SMSlab, blipped-CAIPI) | 1³ | 2×5 | 12 (10×1.2) | 4.2 | 0 (1, +1 AP), 1k (6), 2k (6) | 0.6 | 87/250 | 1.6 | 19.2 | 1 | 4:10 |
|   |   | 8 (2D SMS, blipped-CAIPI) | 1³ | 2×50 | 1 |   | 0 (1, +1 AP), 1k (6), 2k (6) | 0.6 | 87/-- | 9.6 | 9.6 | 2 | 4:10 |
|   |   | 9 (2D) | 1³ | 1×100 | 1 |   | 0 (1, +1 AP), 1k (6), 2k (6) | 0.6 | 87/-- | 19.2 | 19.2 | 1 | 4:10 |
| 4 | 3 | 10 (SMSlab, blipped-CAIPI) | 1³ | 2×7 | 12 (10×1.2) | 5.8 | 0 (1, +1 AP), 1k (32) | 0.6 | 79/245 | 2.1 | 25.6 | 1 | 14:05 |
|   |   | 11 (2D SMS, blipped-CAIPI) | 1³ | 2×70 | 1 |   | 0 (1, +1 AP), 1k (32) | 0.6 | 79/-- | 12.8 | 12.8 | 2 | 14:05 |
|   |   | 12 (2D SMS, blipped-CAIPI) | 1³ | 2×70 | 1 |   | 0 (2, +1 AP), 1k (64) | 0.6 | 79/-- | 12.8 | 12.8 | 1 | 14:05 |

Note:

Abbreviations: Exp. #, experiment NO.; Sub. #, subject NO.; $R_{mb}$, number of simultaneously excited slabs; $N_z$, number of slices per slab, including over-sampled slices; $N_{target}$, number of slices per slab, excluding over-sampled slices; OS, over-sampling rate; $z_{gap}$, center distance between simultaneously excited slabs; $FOV_z$, slab thickness (including over-sampled slices); PF, partial Fourier factor; NSA, number of signal averages. $N_{dir}$: number of diffusion directions.

[a] In SMSlab, each slab was acquired with two over-sampled slices (one at each boundary) to reduce slab boundary artifact.

[b] An additional b=0 scan was conducted with reversed phase encoding gradient (Anterior-Posterior, AP) for susceptibility distortion correction.

[c] Time per volume = TR × $N_z$.

[d] Total scan time = Time per volume × $N_{dir}$ × NSA.

# Supporting Information

### $\varphi_{gap}$ correction

(1) The 1st method: RF phase modulation

For a gradient-echo EPI based blipped-SMSlab sequence, the RF pulse with phase modulation can be expressed as

$$RF_{MB}(t) = \sum_{m=0}^{R_{mb}-1} RF(t) \cdot \exp(i(\omega_m t - \varphi_{gap})) \qquad (1)$$

$RF(t)$ is the basic single-band sub-pulse, $\omega_m$ is the center frequency for the $n_m{}^{th}$ band.

In dMRI that is based on spin-echo EPI, the acquisition of image echo is after two RF pulses: the excitation pulse and the 1st refocusing pulse. Then the requirement of RF modulation is that the net RF encoding phase from the two pulses can cancel $\varphi_{gap}$. For SMSlab dMRI which acquires a 2D navigator, the acquisition of the navigator is after three RF pulses, i.e., one excitation pulse and two refocusing pulses. In this case, the requirement of RF modulation is that the net RF encoding phase from the excitation and two refocusing pulses return to zero, since the 2D navigator has no $k_z$ encoding. A detailed description of how to modify the three RF pulses is based on the reference [1], and the three RF pulses are briefly expressed as follows.

The excitation RF pulse is

$$RF_{MB,ex}(t) = \sum_{m=0}^{R_{mb}-1} RF_{ex}(t) \cdot \exp(i(\omega_m t - \varphi_{gap})) \qquad (2)$$

The first refocusing RF pulse is

$$RF_{MB,echo\_1}(t) = \sum_{m=0}^{R_{mb}-1} RF_{echo\_1}(t) \cdot \exp(i(\omega_m t - \varphi_{gap})) \qquad (3)$$

The second refocusing RF pulse is

$$RF_{MB,echo\_2}(t) = \sum_{m=0}^{R_{mb}-1} RF_{echo\_2}(t) \cdot \exp(i(\omega_m t - \varphi_{gap}/2)) \qquad (4)$$

where $RF_{\text{ex}}(t)$, $RF_{\text{echo\_1}}(t)$ and $RF_{\text{echo\_2}}(t)$ are the basic single-band sub-pulses. In this study, the excitation pulse is applied to +x axis while the two refocusing pulses are applied to +y axis. $RF_{\text{echo\_1}}(t)$ and $RF_{\text{echo\_2}}(t)$ are the same in this study.

Such phase modulation is similar to that adopted in the SMSlab 3D k-space framework [1,2], except that the phase calculation is different. In the 3D k-space framework, $\varphi_{\text{gap,3D}} = 2\pi \cdot d_{\text{gap}}/(R_{mb} \cdot N_z \cdot \Delta z) \cdot n_{kz}$, in which $d_{\text{gap}}$ is the distance between the nearest edges of the simultaneously excited slabs [1,2], as marked on Figure 2A.

(2) The 2nd method: correction during reconstruction

Besides RF phase modulation, $\varphi_{\text{gap}}$ can also be corrected by reconstruction alone. In this method, without RF phase modulation, the acquired data contains the undesired phase $\varphi_{\text{gap}}$. Since $\varphi_{\text{gap}}$ varies among different bands and $k_z$ encodings, it can be treated in a similar way compared with the inter-kz-shot motion-induced phase. Therefore, the reconstruction pipeline for the b=0 s/mm² images can directly follow the DW reconstruction procedure in Figure 4, except that the processing of inter-kz-shot phase should be replaced by $\varphi_{\text{gap}}$ handling. For the DW data, the reconstruction pipeline remains unchanged, except that the processing of inter-kz-shot phase should contain both motion-induced phase and $\varphi_{\text{gap}}$.

Note that if using this $\varphi_{\text{gap}}$ correction method, the SVD-based Nyquist ghost correction algorithm becomes problematic, since the preliminary $\varphi_{\text{ramp}}$ correction (Figure 4A, gray rectangle) cannot be directly conducted in the presence of $\varphi_{\text{gap}}$, which differs across different $k_z$. Therefore, ghost correction based on reference scans can be alternatively considered [3]. In this case, we use the ghost-related phase estimated from the RF modulation acquisitions for Nyquist ghost correction of the data without RF modulation, to test the feasibility of the second $\varphi_{\text{gap}}$ correction method.

# Figures

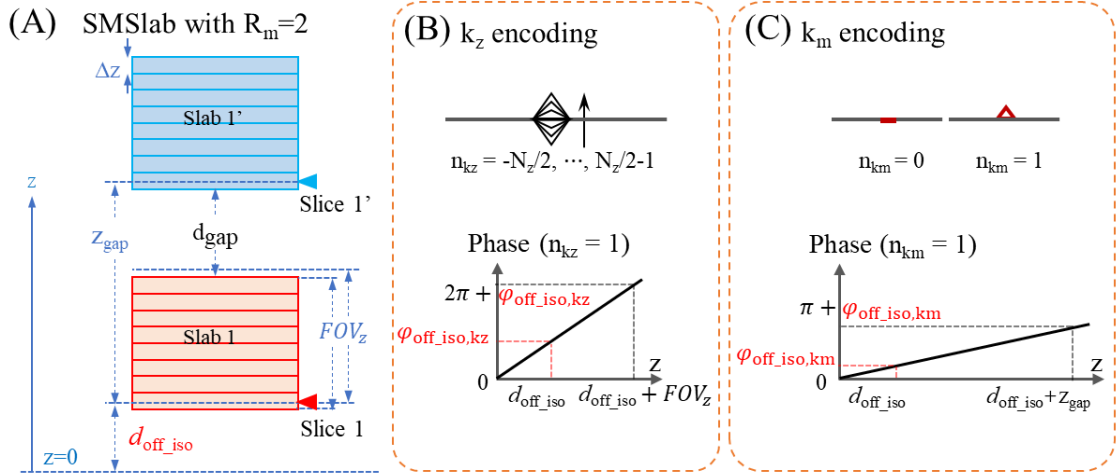

Figure S1. (A) SMSlab excitation ($R_{mb}=2$) when the reference slice (Slice 1) is not at the isocenter ($d_{off\text{-}iso}$ away from z=0). (B) The phase along z axis after $k_z$ encoding ($n_{kz}=1$ for illustration). (C) The phase along z axis after $k_m$ encoding ($n_{km}=1$ for illustration).

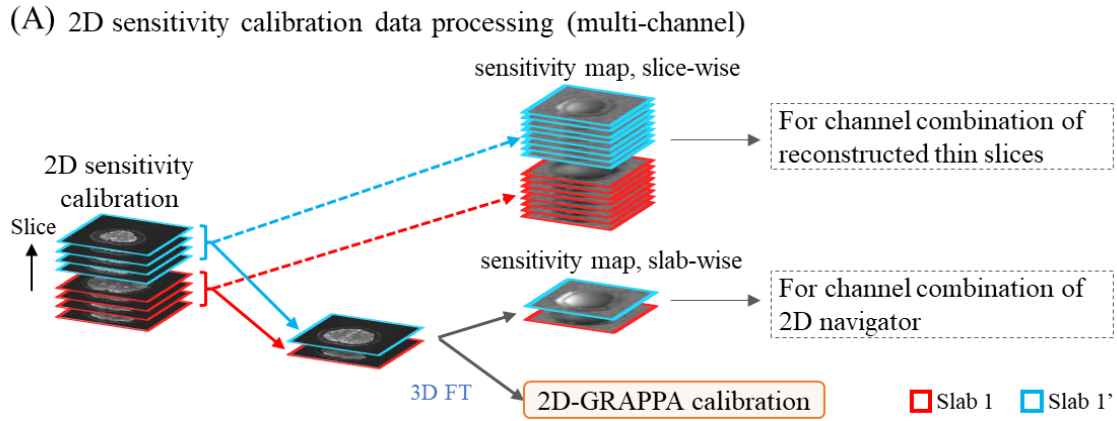

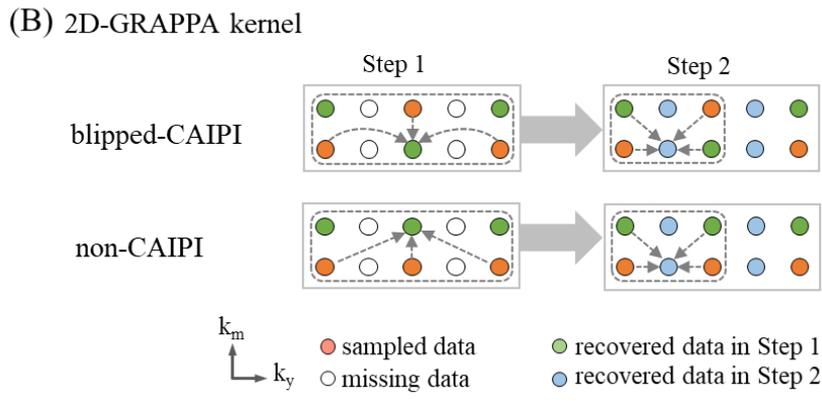

Figure S2. (A) The processing of 2D sensitivity calibration data. The 2D sensitivity calibration data were used to calculate the sensitivity maps for channel combination of reconstructed thin slices and the 2D navigator, as well as synthesization of the 2D-GRAPPA calibration data. These data all contain multi-channel information, yet omitted in the figure for simplicity. (B) The 2D-GRAPPA interpolation kernel for SMSlab with the blipped-CAIPI and non-CAIPI sampling pattern, respectively. The two sampling patterns use the same interpolation kernel size, but they have different interpolation ways.

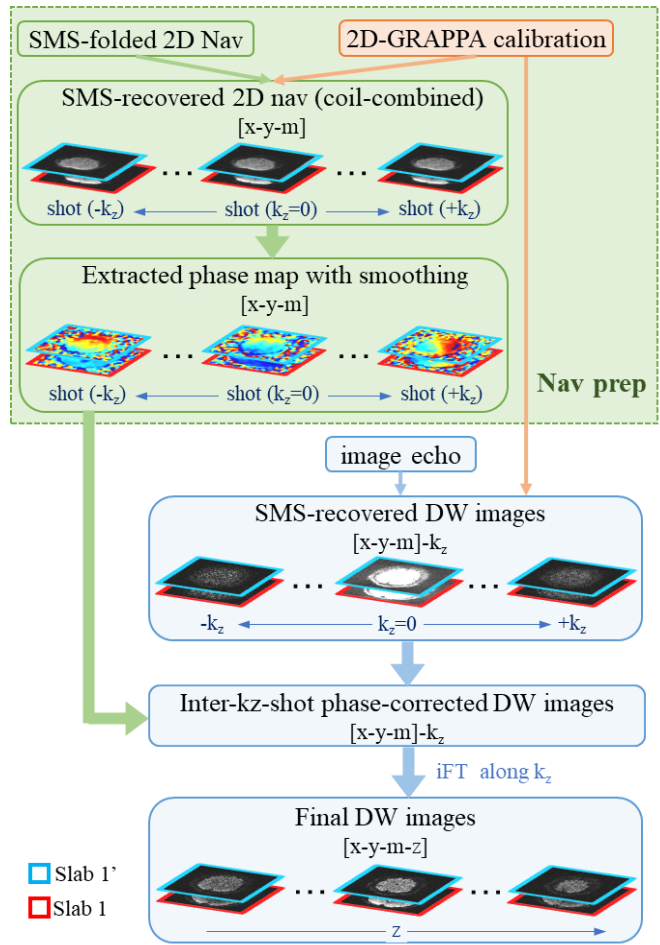

Figure S3. Inter-kz-shot motion-induced phase correction of the diffusion-weighted (DW) data.

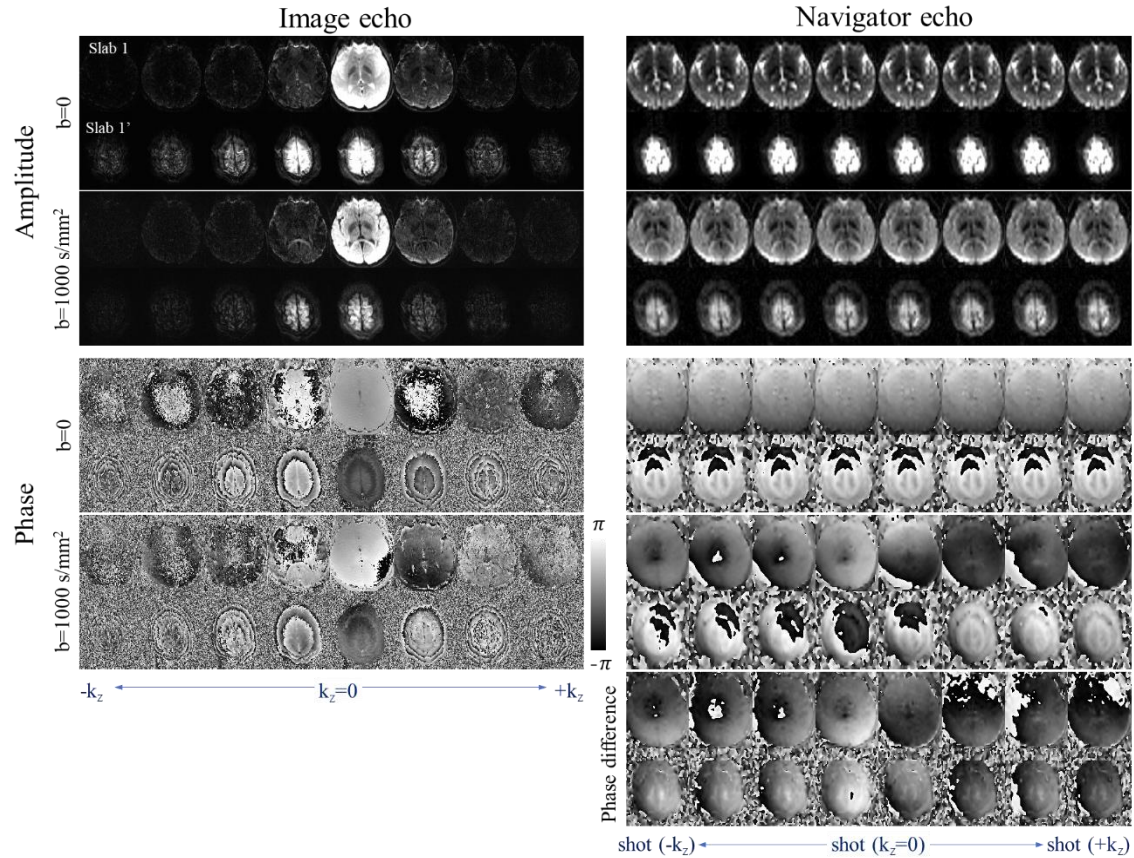

Figure S4. Example interim magnitude and phase images of the image echo and navigator echo. The phase maps of the image echo are without inter-kz-shot phase correction. The phase maps of the b=0 image echo also varies across different $k_z$ shots due to different $k_z$ encoding gradients. The phase difference maps in the right panel shows the phase difference between the b=1000 s/mm² and b=0 s/mm² navigator images. Note that besides the phase difference maps, the phase maps of the diffusion-weighted (DW) navigators alone can also be used for inter-kz-shot phase correction. The rationality lies in that the 2D DW navigator records both background phase and diffusion-related phase, of which the diffusion-related phase varies among different $k_z$ shots, while the background phase keeps consistent across all $k_z$ shots.

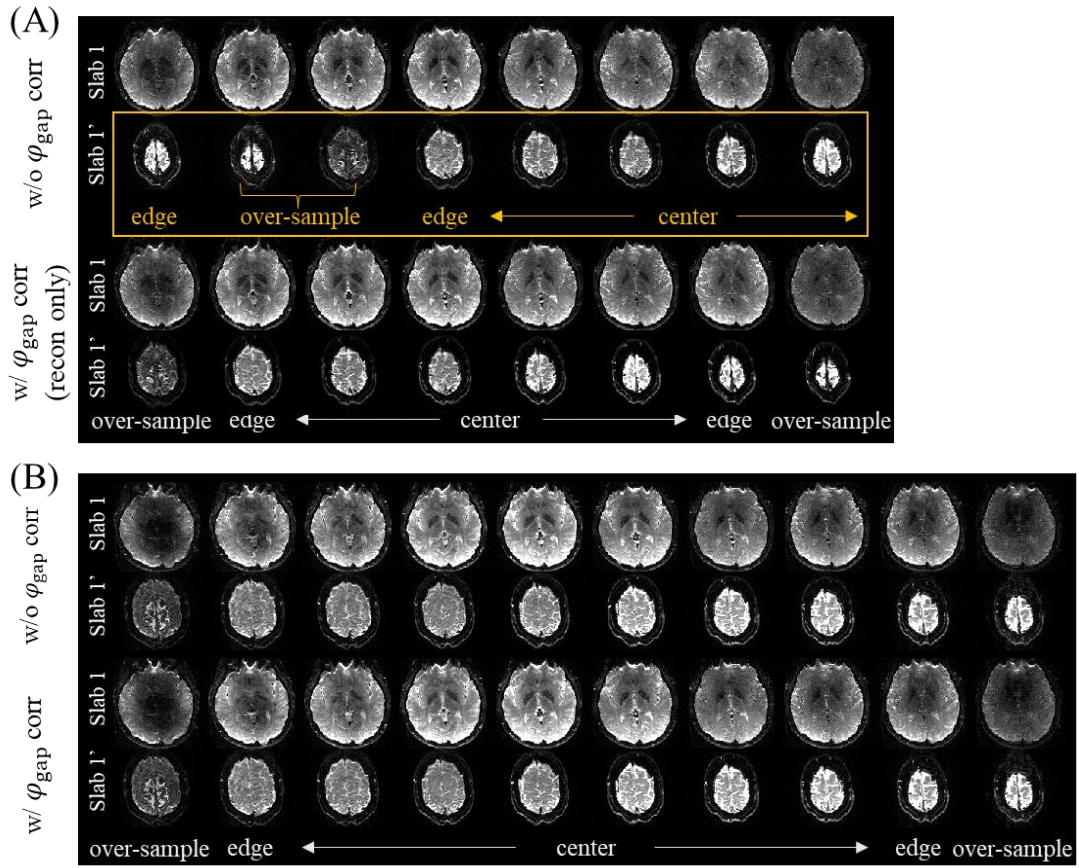

Figure S5. (A) When the ratio between the inter-slab center distance ($z_{\text{gap}}$) and $FOV_z$ is not an integer (scan 1 in Table 1, $z_{\text{gap}}$ modulo $FOV_z$ equals to 2), image shift occurs in Slab 1' when without $\varphi_{\text{gap}}$ correction. When not using RF modulation during acquisition, the $\varphi_{\text{gap}}$-induced problem can be addressed by correction during reconstruction alone (the bottom panel in A). (B) When the ratio $z_{\text{gap}}/FOV_z$ is an integer (scan 2 in Table 1), blipped-SMSlab imaging results are the same with or without $\varphi_{\text{gap}}$ correction. The b=0 s/mm$^2$ images are shown for illustration.

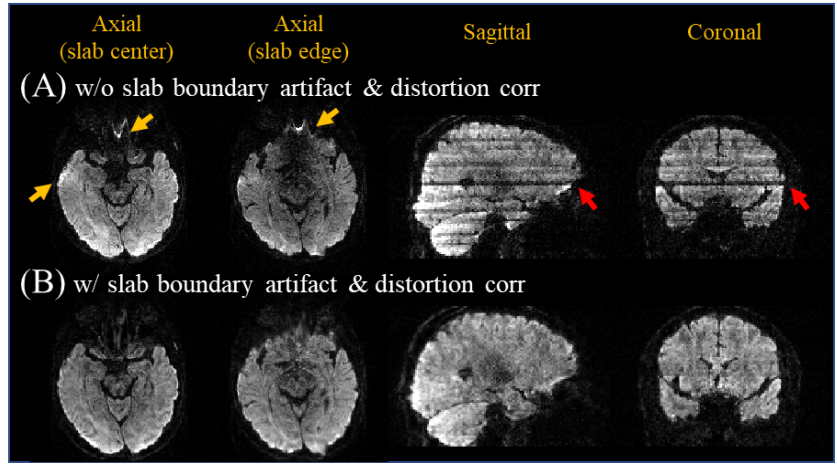

Figure S6. The blipped-SMSlab diffusion images before slab boundary artifacts and distortion correction (A), as well as the images after correction (B). Two axial slices (acquisition view), including one at slab center and the other at slab edge, one sagittal slice and one coronal slice are displayed. Image distortion and slab boundary artifacts are indicated by the yellow and red arrows, respectively.

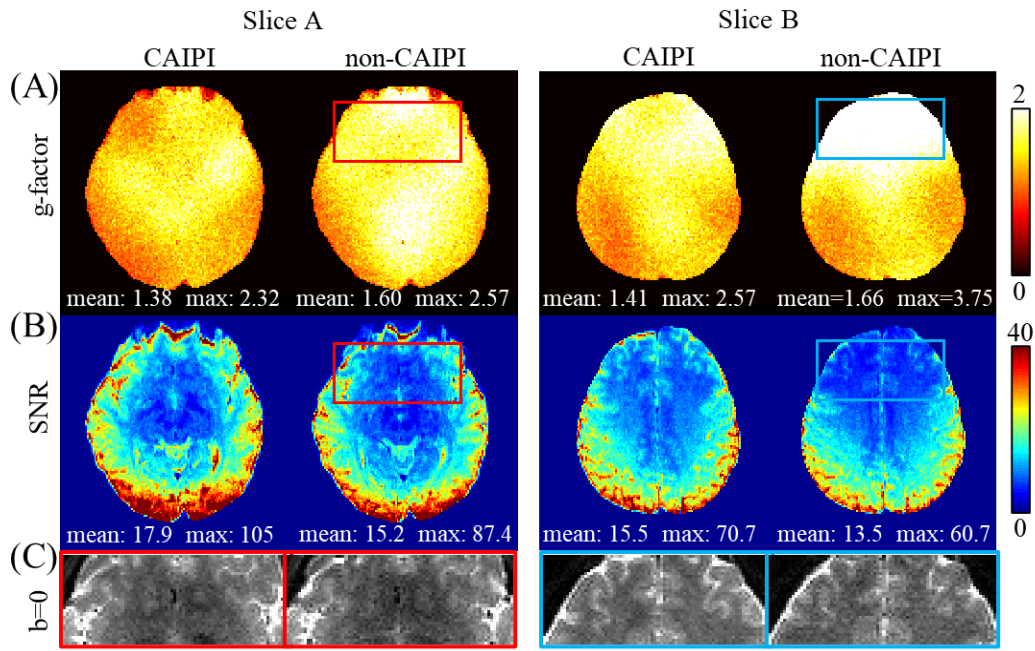

Figure S7. SMSlab imaging with the blipped-CAIPI or non-CAIPI sampling patterns. (A) g-factor. Higher g-factor means more g-factor related SNR loss. (B) SNR map. (C) Zoomed-in areas from the b=0 images.

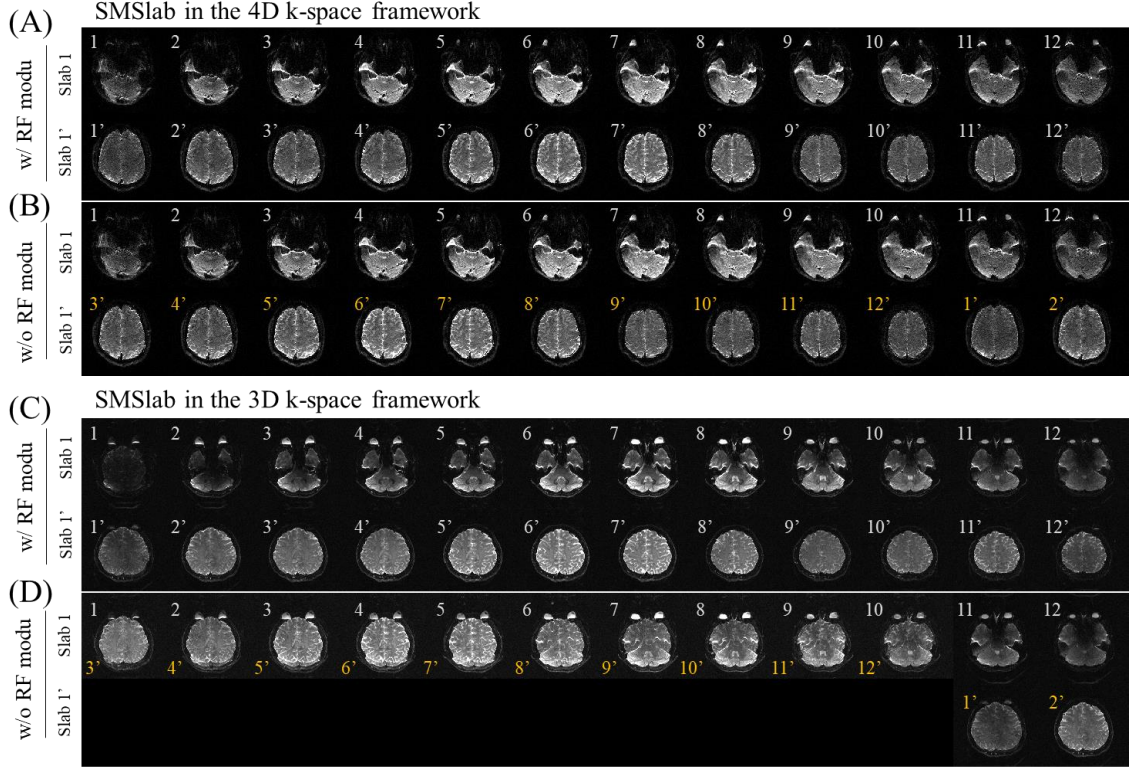

Figure S8. SMSlab images in the 4D and 3D k-space frameworks. In the 4D k-space framework (A-B), when the ratio between the inter-slab center distance $z_{\text{gap}}$ and $FOV_z$ is not an integer number, the images in Slab 1' shift along the intra-slab dimension without RF phase modulation (B), and the shift distance is equal to the slice number of $z_{\text{gap}}$ modulo $FOV_z$ (10 in this case). Since the images in the two slabs do not interact with each other, this offers the flexibility to use RF phase modulation or use correction during reconstruction alone. While in the 3D k-space framework (C-D), when the ratio between the inter-slab gap $d_{\text{gap}}$ and the extended $FOV_z$ by concatenating the simultaneously excited slabs, i.e., $d_{\text{gap}}/(R_{mb} \cdot N_z \cdot \Delta z)$, is not an integer number, the images in Slab 1' also shifted without RF phase modulation (D), but they shifted within the extended $FOV_z$. The shift distance equals to the result of $d_{\text{gap}}$ modulo $R_{mb} \cdot N_z \cdot \Delta z$ (10 in this case). As a consequence, the shifted images alias together with the images in Slab 1, which cannot be shifted back, and RF phase modulation is needed when using k-space-based reconstruction methods.

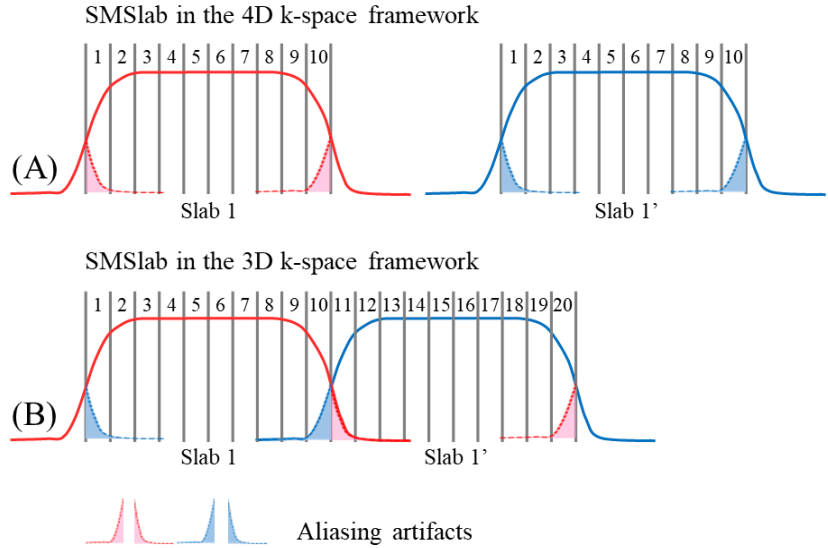

Figure S9. The aliasing pattern of SMSlab in the 4D k-space framework (A) and the 3D k-space framework (B), respectively. In the 4D k-space framework, the signals excited out of the defined slabs manifest as intra-slab aliasing, which is the same with the single-band 3D multi-slab imaging. In the 3D k-space framework, the excited signals out of the defined slabs alias with the edge slice at the opposite side of the other sub-slab. Such a difference should be noticed when modeling the slab boundary artifacts during the correction.

(A) 2D SMS-EPI, $N_{dir}$ =32, NSA=2

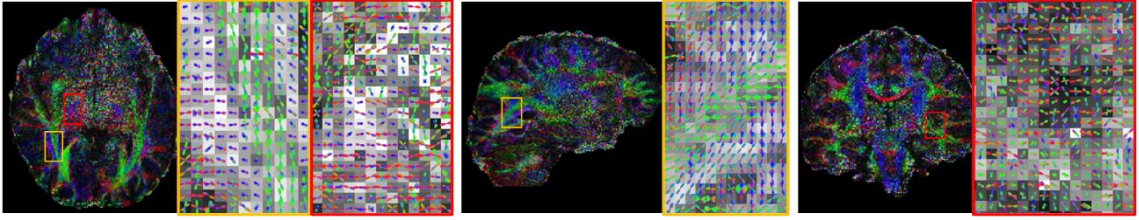

(B) 2D SMS-EPI, $N_{dir}$ =64, NSA=1

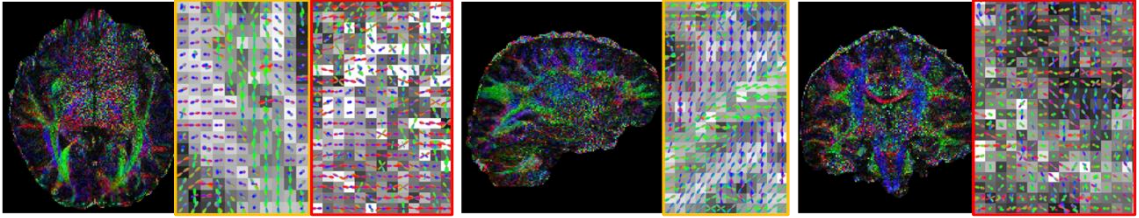

(C) blipped-SMSlab, $N_{dir}$ =32, NSA=1

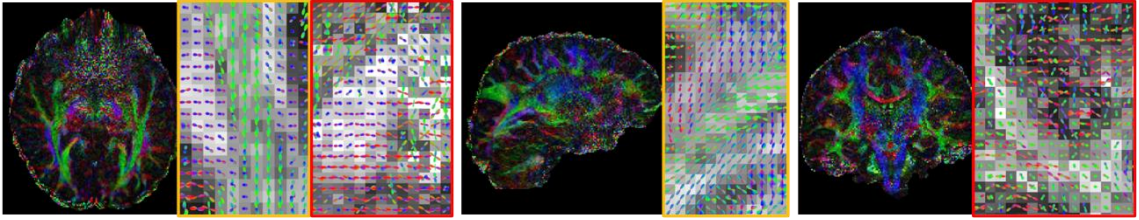

Figure S10. Comparison of color-coded FA (cFA) maps and fiber orientation distribution function (ODF) between blipped-SMSlab and 2D SMS-EPI dMRI at 1.0 mm isotropic resolution. These data used MPPCA denoising [4]. (A) 2D SMS-EPI dMRI with 32 diffusion directions and 2 averages. (B) 2D SMS-EPI dMRI with 64 diffusion directions and 1 average. (C) The blipped-SMSlab dMRI with 32 diffusion directions and 1 average. The acquisition time was the same for the three acquisitions (~14 min for each). $N_{dir}$: number of diffusion directions.

**Blurring effect in 2D and 3D imaging along the slice direction**

For the x and y dimensions, the blurring effect is expected to be the same for the 2D and 3D encoding methods when using the same readout configurations of the Gx and Gy gradients. As for the slice direction, their blurring effects are affected in different aspects.

For the 2D methods, the non-ideal slice profile excites signals of the neighboring slices (Figure S11A), which blurs the images along the slice direction after image reconstruction. In actual imaging, the slice profile may be further distorted by several factors, such as hardware imperfections, $B_0$ and $B_1$ field inhomogeneities.

For the 3D multi-slab and SMSlab methods, the non-ideal slab profile also excites signals of the neighboring slabs, but this results in aliasing instead of blurring along the slice direction because of $k_z$ encoding. With an actual slab profile, the aliasing and signal modulation problem can be well addressed after slab boundary artifact correction. However, the slab profile estimated from the calibration scan cannot be the same with the actual slab profile due to inter-scan motion and reconstruction artifacts. As a consequence, the slab boundary artifact correction process can introduce the blurring effect to some extent, which is related to the regularization term or loss function used to suppress boundary artifacts, as evaluated in the previous work of CPEN [5]. The regularization term or loss function can be fine-tuned to reduce the blurring effect while maintaining satisfactory correction performance [5].

For Fourier encoding and reconstruction, (1) the truncation of the Fourier series, i.e., only finite k-space encodings are implemented, as well as (2) the discrete encoding and sampling, result in the point spread function (PSF) below [6,7]:

$$h(x) = N \cdot \Delta k \frac{sinc(\pi \cdot N \cdot \Delta k \cdot x)}{sinc(\pi \cdot \Delta k \cdot x)} e^{-i\pi \cdot \Delta k \cdot x}$$

For 3D-based imaging, when the number of $k_z$ gradient encodings is N = 12, the PSF and magnitude PSF along the slice direction are shown in Figures S11B and S11C, respectively. The effective width of h is half the width of its main lobe, which is the target resolution [6,7]. Therefore, the voxels are well defined in 3D-based imaging. The truncation of the Fourier series causes Gibbs ringing (also known as truncation artifacts).

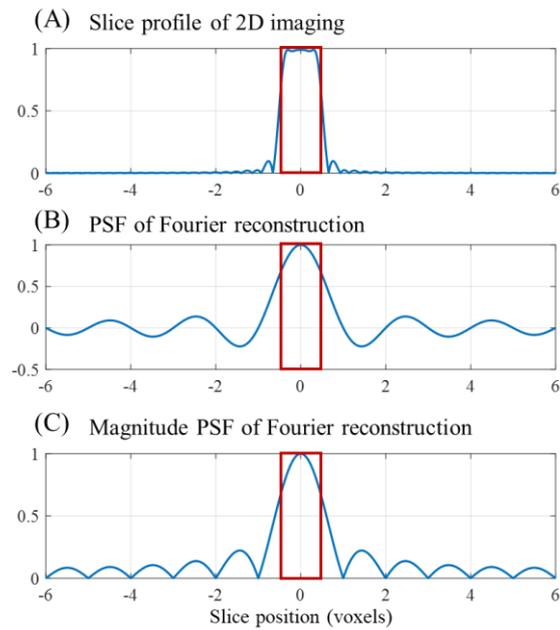

Figure S11. (A) The RF slice profile from Bloch simulation for 2D imaging (single band for illustration). This RF pulse uses a default waveform on our scanner, with a time-bandwidth produce of 4. (B) The PSF of 3D Fourier reconstruction along the slice direction, when the total number of $k_z$ gradient encodings is N = 12. (C) The magnitude of the PSF shown in panel (B).

**Tables**

Table S1. Detailed imaging parameters of the calibration scans for imaging scans in Table 1.

| | Exp.# | Scan # (type) | Res. (mm$^3$) | $R_{mb}$ ×Slabs | $N_{shot-ky}$ | $N_z$ ($N_{target}$×OS)[a] | PF | TE1 (ms) | TR (s) | Total scan time (min:s)[b] |
|---|---|---|---|---|---|---|---|---|---|---|
| Slab profile calibration for SMSlab | 1 | 1 (blipped-CAIPI) | 1.5$^3$ | 2×7 | 1 | 16 (6×2.67) | 0.7 | 77 | 1.8 | 0:29 |
| | | 2 (blipped-CAIPI) | 1.5$^3$ | 2×5 | 1 | 20 (8×2.5) | 0.7 | 77 | 1.8 | 0:36 |
| | 2 | 3 (non-CAIPI) | 1.5$^3$ | 2×5 | 1 | 20 (8×2.5) | 0.7 | 77 | 1.8 | 0:36 |
| | 3 | 4 (blipped-CAIPI) | 1.3$^3$ | 2×7 | 1 | 20 (8×2.5) | 0.65 | 85 | 1.8 | 0:36 |
| | | 7 (blipped-CAIPI) | 1$^3$ | 2×5 | 1 | 24 (10×2.4) | 0.6 | 87 | 1.6 | 0:38 |
| | 4 | 10 (blipped-CAIPI) | 1$^3$ | 2×7 | 1 | 24 (10×2.4) | 0.6 | 79 | 2.1 | 0:51 |
| 2D sensitivity calibration | 1, 2 | 1~3 (2-shot EPI) | 1.5×1.5×3 | 1×52 | 2 | 1 | 1 | 79 | 7 | 0:14 |
| | 3 | 4~6 (2-shot EPI) | 1.3×1.3×2.6 | 1×60 | 2 | 1 | 1 | 98 | 10 | 0:20 |
| | 3, 4 | 7~12 (2-shot EPI) | 1×1×2 | 1×78 | 2 | 1 | 0.7 | 66 | 12 | 0:24 |

Note:

Abbreviations: Exp. #, experiment NO.; $R_{mb}$, number of simultaneously excited slabs; $N_z$, number of slices per slab, including over-sampled slices; $N_{target}$, number of slices per slab, excluding over-sampled slices; OS, over-sampling rate; PF, partial Fourier factor; NSA, number of signal averages.

[a] In SMSlab, each slab was acquired with two over-sampled slices (one at each boundary) to reduce slab boundary artifacts.

[b] Total scan time = TR×$N_{shot-ky}$×$N_z$